\documentclass[%
 reprint,
superscriptaddress,
 amsmath,amssymb,
 aps,
 showkeys,
 nolongbibliography,
prb,
floatfix
]{revtex4-2}

\usepackage{longtable}
\usepackage{adjustbox}
\usepackage [utf8]{inputenc}
\usepackage{dcolumn}% Align table columns on decimal point
\usepackage{bm}% bold math
\usepackage{float}
\usepackage{lipsum, babel}
\usepackage{import}

\begin{document}

\preprint{APS/123-QED}

\title{$\mathrm{C}_{60}$+$\mathrm{C}_{60}$ molecular bonding revisited and expanded}

\author{Jorge Laranjeira}
\email{jorgelaranjeira@ua.pt}
\affiliation{Departamento de Física and CICECO, Universidade de Aveiro, 3810-193 Aveiro, Portugal}

\author{ Karol Struty\'{n}ski}
\affiliation{Departamento de Química and CICECO, Universidade de Aveiro, 3810-193 Aveiro, Portugal}
\author{ Leonel Marques}
\affiliation{Departamento de Física and CICECO, Universidade de Aveiro, 3810-193 Aveiro, Portugal}
\author{ Emilio Martínez-Núñez}
\affiliation{Departamento de Química Física, Universidade de Santiago de Compostela, 15782, Santiago de Compostela, Spain}
\author{ Manuel Melle-Franco}
\email{manuelmelle@ua.pt}
\affiliation{Departamento de Química and CICECO, Universidade de Aveiro, 3810-193 Aveiro, Portugal}

%%%%%%%%%%%%%%%%%%%%%%%%%%%%%%%%%%%%%%%%%%%%%%%%%%%%%%%%%%%%%%%%%%%%%
%% The document title should be given as usual. Some journals require
%% a running title from the author: this should be supplied as an
%% optional argument to \title.
%%%%%%%%%%%%%%%%%%%%%%%%%%%%%%%%%%%%%%%%%%%%%%%%%%%%%%%%%%%%%%%%%%%%

\date{\today}% It is always \today, today,

\begin{abstract}
Several dimerization products of fullerene $\mathrm{C}_{60}$ are presented and thoroughly characterized with a quantum chemical DFT model augmented by dispersion. We reanalyze and expand significantly the number of known dimers from 12 to 41. Many of the novel bonding schemes were found by analyzing more than 2 nanoseconds of high energy molecular dynamics semiempirical trajectories with AutoMeKin, a methodology previously used to compute the reactivity of much smaller molecules. For completeness, this was supplemented by structures built by different geometric considerations. Also, spin-polarization was explicitly considered yielding 12 new bonding schemes with magnetic ground states. The results are comprehensively analyzed and discussed in the context of yet to be explained 3D fullerene structures and recent fullerene 2D systems.
\end{abstract}

 
\maketitle

\section*{Introduction}
\begin{figure}[H]
	%	\hspace*{-1.5cm} %shiftar à esquerda
	\centering
	\includegraphics[scale=1.5]{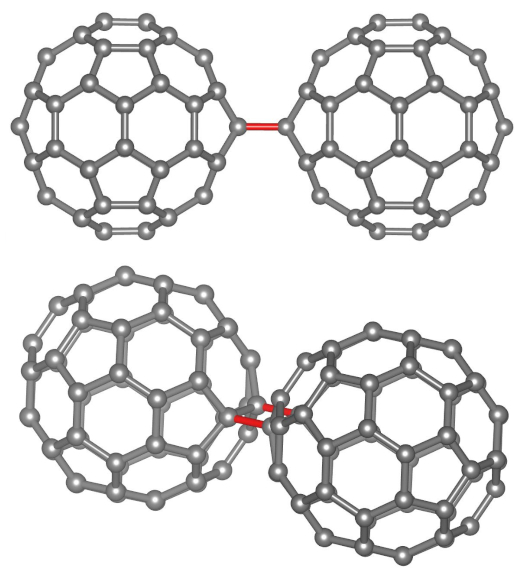}
	\caption{Two views of the most stable and well studied $\mathrm{C}_{60}$ dimer with molecules bonded via 66/66 2+2 cycloaddition.}
 \label{fig0}
\end{figure}

At room temperature and atmospheric pressure, $\mathrm{C}_{60}$ is a van der Waals solid with a face-centered cubic (fcc) structure where the molecules are rotating freely \cite{i21}. When subjected to visible or ultraviolet light, or high-pressures high-temperatures (HPHT) treatments the $\mathrm{C}_{60}$ molecules bond to each other via 66/66 2+2 cycloaddition forming aggregates and dimers \cite{i18,i57,strout1993dim}, see figure \ref{fig0}.

Increasing the pressure and temperature of the $\mathrm{C}_{60}$ HPHT treatment leads to the formation of extended crystalline networks \cite{i17,i6,laran2017,laran2018,i12,i68,LARANJEIRA2022}. At pressures below 8 GPa the formation of low-dimensionality polymers occurs yielding 1D orthorhombic, 2D tetragonal and 2D rhombohedral phases. In all these low-dimensionality polymers, the molecules arrange themselves in covalently bonded chains (1D) or sheets (2D) formed exclusively by 66/66 2+2 cycloadditions. 

It is above 8 GPa that fully 3D polymerized crystals are formed \cite{laran2018,i12,i68,LARANJEIRA2022}. Here other bonding schemes, besides the 66/66 2+2 cycloaddition, start to play an important role, in fact, 3D polymers proposed to date have different bonding schemes such as the 56/56 2+2 cycloaddition \cite{laran2017,laran2018}, the 6/6 3+3 cycloaddition and the double 66/66 4+4 cycloaddition \cite{i12}, the 56/65 2+2 cycloaddition and the 5/5 3+3 cycloaddition \cite{i68} or even the double 5/5 2+3  cycloaddition \cite{LARANJEIRA2022}.

Recently, other $\mathrm{C}_{60}$ polymers have been produced from a different route \cite{naturemgc60_22,naturemgc60_23}. This has yielded a pure carbon bulk material based on covalently linked fullerenes forming 2D sheets with hexagonal symmetry dubbed graphurelene \cite{naturemgc60_23}. These layers are bonded via single-bond and 56/65 2+2 cycloaddition, unlike any 2D polymers synthesized to date by HPHT treatment, showcasing the potential of alternative synthetic routes to produce new structures. 

%2D hexagonal sheets of polymerized $\mathrm{C}_{60}$,  have been produced based on an ionic material where the counter-ion is either retired as catalyst to obtain a bulk, layered, crystal, similar to the 2D phases of HPHT synthesized polymers, dubbed graphulerite, that was then exfoliated to obtain the 2D sheets, graphulerene \cite{naturemgc60_22,naturemgc60_23}. This layers are bond via single bond and 56/65 2+2 cycloaddition unlike the previous 2D polymers.***

3D $\mathrm{C}_{60}$ polymers show quite remarkable physical properties for pure carbon phases, such as metallicity and low-compressibility which fosters interest in these materials \cite{laran2018,i12,i68,LARANJEIRA2022}. Nevertheless, diffraction patterns from these phases lack the resolution needed to solve their crystalline structure. Thus, it is necessary to build complementary computational models fitting the overall symmetry and lattice parameters to the diffraction pattern and study their stability recurring to quantum simulation tools like Density Functional Theory (DFT) \cite{laran2017}.

The distances between the two $\mathrm{C}_{60}$ covalently bonded molecules, i.e. a $\mathrm{C}_{60}$ dimer, computed from their molecular centers are similar to the corresponding distance in a bulk crystalline phase. For instance, the $\mathrm{C}_{60}$ molecular distance, we compute for the 66/66 2+2 cycloaddition dimer, is $\sim$9.1 \AA \space while in the low-dimensionality polymers it is found to be between 9.02 and 9.18 \AA \space \cite{i17,i6}. Also, for the recent graphulerene  \cite{naturemgc60_22,naturemgc60_23}, the computer dimer distances are 9.23 and 9.09 \AA \space while the experimental bonding distance of 9.23 and 9.16 \AA \space respectively. Hence, relating bonding distances to bonding schemes constitutes a simple yet accurate interpretation tool for the $\mathrm{C}_{60}$ crystal structure assignment, rendering covalent $\mathrm{C}_{60}$ dimers specifically relevant to compute and understand. 

%***Another recent example are the single bond and 56/65 2+2 cycloaddition bonds distances found in 2D hexagonal sheets of polymerized $\mathrm{C}_{60}$ dubbed graphulerene \cite{naturemgc60_22,naturemgc60_23}, which are in striking agreement with our result. In particular, the experimental bonding distances are 9.23 and 9.16 \AA, for the single bond and 56/65 2+2 cycloaddition, repectively, while the dimer ones are 9.23 and 9.09 \AA.*** 

%Thus, it is relevant to explore possible $\mathrm{C}_{60}$ dimers as their bonding schemes may help to understand better $\mathrm{C}_{60}$ crystalline phases. In fact, relating bonding distances to bonding schemes constitutes a simple yet powerful interpretation tool for the $\mathrm{C}_{60}$ crystal structure modeling and crystal structure assignment. 

In the 90's and beginning of the 2000's, several studies regarding dimers were presented, nevertheless, to the best of our knowledge, no more than 12 dimers have been described altogether. Here, We consider, unless noted,  $\mathrm{C}_{60}$ covalent dimers, that is dimers where the forming fullerene molecules are joined by intermolecular bonds. Strout et al. \cite{strout1993dim} performed probably the most extensive $\mathrm{C}_{60}$ dimerization study which addressed the 2+2 cycloadditions and a 2+4 cycloaddition. This last bonding scheme, following our notation, corresponds to a double 6/56 4+2 cycloaddition, since two 2+4 cycloaddition reactions, involving one hexagon (6) and one intramolecular single bond (56) each, are present. 
%Not sure this relevant
%In addition, besides bonding between $\mathrm{C}_{60}$ molecules, they also studied molecular structures obtained by $\mathrm{C}_{60}$ coalescence.

Later, in 1994, Adams et al. \cite{adamsprb94_dimeros} proposed three other bonding schemes, two bonding the molecules via their hexagons and one between their pentagons. In the bonding between hexagons, the six atoms from one hexagon bond to the six atoms of the other hexagon, which can be obtained in the following two ways: 1) the hexagons of the two molecules align with each other (as well as the pentagons); 2) the hexagons from one molecule align with the pentagons of the other. When possible, we will keep the cycloaddition notation for the formed dimers nomenclature thus, in this case the dimers will be named, for instance, "triple 66/66 2+2 cycloaddition" (or "triple 56/56 2+2 cycloaddition") and "triple 56/66 2+2 cycloaddition" respectively. In the bonding between pentagons the five atoms from one pentagon bond to the five atoms of the other, the dimer will be referred to as "double 56/56 2+2 cycloaddition plus single bond". More details about this bonding nomenclature are available in the supporting information (SI) section A. In 2007 Liu et al. \cite{LIU2007_dimeros} studied again the two dimers bonded via hexagons confirming that they are thermodynamically stable and thus, could be observed experimentally.   

Osawa et al. \cite{osawa_dimeros} reported the two possible 6/6 4+4 cycloaddition bonded dimers, while Matsuzawa et al. \cite{Matsuzawa_dimeros} also considered the 6/6 4+4 cycloaddition dimers and added one of the two 66/6 2+4 cycloaddition dimers to their calculations. Besides the mentioned dimers, the single bond dimer (SB) was also considered \cite{i26}. In all of the works mentioned so far, the lowest energy dimer was consistently found to be the 66/66 2+2 cycloaddition \cite{strout1993dim,i26,LIU2007_dimeros,adamsprb94_dimeros}. % but as these dimmers where a lot less stable thant the 66/66 2+2 ones they

Apart from dimers formed just with pristine, i.e. unfunctionalized, $\mathrm{C}_{60}$, there have also been numerous studies of dimers formed with other intermediates as well as coalescence products \cite{fowler_dimeros,strout1993dim,diudea_dimeros}. In addition, there have been studies on the dimerization and coalescence of $\mathrm{C}_{60}$ inside carbon nanotubes \cite{i17,Koshino_dimeros,2022encap}. Note that although some of these systems also did appear in our calculations, for simplicity, they were not considered in our analysis and will not be discussed. Nevertheless, coalescence products might improve the understanding of amorphous carbon phase formation from $\mathrm{C}_{60}$ \cite{soldatov2020} or other phases were the $\mathrm{C}_{60}$ cage is broken \cite{pan2023long} and might be the object of future studies.   

The new systems reported here have been obtained fundamentally with AutoMeKin \cite{amk1,amk2,amk3}, an automatic computational procedure to find and characterize chemical reaction paths, and extended by manual construction following geometric considerations. We report a total of 41 stable dimers computed at DFT level. From these, 25 were obtained with AutoMeKin, the others were built from geometric considerations or obtained from the literature. The molecular structure and relative stability of all dimers were computed at the TPSS-Def2-TZVPP/B3LYP-6-31G(d,p) theory level augmented by D3 dispersion with Becke–Johnson damping. All the dimers studied present $\mathrm{C}_{60}$ distances between 8.04 \AA \space and 9.23 \AA \space which renders some of these bridging patterns suitable candidates to be present in 3D $\mathrm{C}_{60}$ polymers. 

\section*{Methods} 

First, we performed a benchmark calculation of the dimer bonding energies for four different Hamiltonians, namely: TPSS-Def2-TZVPP-D3BJ/TPSS-6-31G(d,p)-D3BJ, TPSS-Def2-TZVPP-D3BJ/B3LYP-6-31G(d,p)-D3BJ, TPSS-6-31G(d,p)-D3BJ and B3LYP-6-31G(d,p)-D3BJ, on the lowest five energy $\mathrm{C}_{60}$ dimers, table \ref{tcomp_tpssb3lyp}. The TPSS-Def2-TZVPP-D3BJ Hamiltonian was chosen after a recent benchmark study comparing the $\mathrm{C}_{60}$ isomerization energies with 115 methodologies which recommended this method based on its accuracy \cite{comparacaodftc60}. 
The bonding energy was computed by subtracting the energy of two isolated $\mathrm{C}_{60}$ to the energy of each dimer: $\Delta E_{Bond}=E_{dimer}-2E_{\mathrm{C}_{60}}$.
\setlength{\LTcapwidth}{\textwidth}
 \begin{small} 
 \begin{longtable*}{c | c | c | c | c | c }
 %\diag{0em}{1.7cm}{functional}{minima}  & Min1   & Min2   & Min3   & Min4   & Min5   \\
 Theory level  & Min1   & Min2   & Min3   & Min4   & Min5   \\
 	\hline
B3LYP-6-31G(d,p)-D3BJ   &  -0.6541 &   0.2055 &  0.5839  &  1.0571 & 1.0934    \\
 	\hline
TPSS-6-31G(d,p)-D3BJ    &  -0.5415 &   0.2125 &  0.5304  &  0.9635 & 0.9904    \\
 	\hline
 TPSS-Def2-TZVPP-D3BJ/TPSS-6-31G(d,p)-D3BJ    &  -0.2512 &   0.4931 &  0.7398  &  1.2428 & 1.2676    \\
	\hline
 TPSS-Def2-TZVPP-D3BJ/B3LYP-6-31G(d,p)-D3BJ   &  -0.2494 &   0.4989 &  0.7440  &  1.2501 & 1.2771    \\

\caption{Bonding energy of the five lowest energy $\mathrm{C}_{60}$+$\mathrm{C}_{60}$ dimers optimized with different levels of theory given in eV/dimer.}
 	\label{tcomp_tpssb3lyp}
 \end{longtable*}   
 	\end{small}

  \begin{small}  
\begin{longtable*}{c | c | c }
 %\diag{0em}{1.7cm}{functional}{minima}  & Min1   & Min2   & Min3   & Min4   & Min5   \\
 Theory level  & $\Delta \mathrm{E}_{Bond} $ (eV/dimer) & $\Delta \mathrm{E}_{Bond} $+ BSSE (eV/dimer)   \\
 	\hline
{B3LYP-6-31G(d,p)-D3BJ}    & -0.4569 & -0.3512 \\
{TPSS-Def2-TZVPP-D3BJ/B3LYP-6-31G(d,p)-D3BJ} & -0.3604 & -0.3460 \\

\hline

\caption{Bonding energy of the non-covalent $\mathrm{C}_{60}$+$\mathrm{C}_{60}$ dimer optimized with different levels of theory and corrected by the basis set superposition error (BSSE), given in eV/dimer.}
 	\label{bsse}
 \end{longtable*}  
  \end{small} 

  The Basis Set Superposition Error (BSSE) was evaluated in a non-covalent dimer, and found to be sizable, around 0.1 eV, for bonding energies with the smaller, DZP, basis sets, table \ref{bsse}. After the BSSE correction, both levels of theory yielded values differing less than 5 meV. Interestingly, relative energies were very similar, and also similar results were found for all DFT functionals. Considering this, we selected the TPSS-Def2-TZVPP-D3BJ/B3LYP-6-31G(d,p)-D3BJ methodology as it yields a good compromise between accuracy and computational cost. Namely, we performed the geometry optimizations at the B3LYP-6-31G(d,p) \cite{B3lyp1,B3lyp2} level augmented by the D3 van der Waals correction \cite{d3grimmes,d3grimme2} with Becke–Johnson damping \cite{bj} (B3LYP-6-31G(d,p)-D3BJ) followed by a single point calculation with TPSS-Def2-TZVPP Hamiltonian \cite{tpss,def} also augmented by the D3 van der Waals correction with Becke–Johnson damping (TPSS-Def2-TZVPP-D3BJ) for the energies. Hessian calculations were performed in all presented geometries at the B3LYP-6-31G(d,p)-D3BJ level to explicitly check the absence of negative frequencies. All calculations were performed with Gaussian09 \cite{g09}.

We explicitly analyzed the dispersion contribution to the bonding energy, as this contribution is missing in most early DFT and semiempirical studies. The dispersion correction shifts bonding energies to more binding values, for the B3LYP-6-31G(d,p) functional the maximum shift is 1.34 eV and the average one is 0.78 eV. For the TPSS-Def2-TZVPP functional the stabilization is lower, the maximum shift is 1.15 eV and the average 0.68 eV. This indicates that there is a consistent attractive dispersion contribution to the bonding interaction in all covalently bonded $\mathrm{C}_{60}$ dimers. In fact,  without van der Waals correction, all dimers are metastable compared to the isolated $\mathrm{C}_{\mathrm{60}}$ molecule, see SI table S1 and figures S1 and S2. 

 To systematically explore possible minima we employed AutoMeKin (AMK) \cite{amk1,amk2,amk3}, a software designed to find reaction mechanisms. AMK analyzes high-energy molecular dynamics trajectories via graph theory to search for structural transformations. When a structural transformation, such as bonds breaking or forming, is detected it finds and characterizes the transition state via its second-order derivatives. From the transition state, the intrinsic reaction coordinate (IRC) method is then used to obtain reactants and products. For this, 2.25 ns were run altogether via 4500, 500 fs long, independent molecular dynamics of two $\mathrm{C}_{60}$ molecules in a high energy bonding configuration. The dynamics were run with a specially modified version of MOPAC2016 \cite{stewart_2016} using the PM7 \cite{PM7} Hamiltonian 
 with a time step of 0.5 fs. Note that AMK so far has been used in molecules  $\leq 30$ atoms \cite{big_amk}, so this study, on a system with 120 atoms, 
 showcases how the AMK methodology can be successfully extended to other problems. 
 
 % to date by far, with the previous record belonging to ozonolysis of $\alpha$-pinene molecules, having only 30 atoms \cite{big_amk} showcases .

 In addition, for completeness, we have comprehensively searched for reported dimers in literature and also constructed other bonding schemes considering the geometric possibilities allowed by the molecule. In the tables, the following labels reveal the procedure followed for each minimum:  L, literature, G, geometric considerations and AMK. To build dimers from geometric considerations, we went through the possibilities of bonding between each molecular element of geometry avoiding large deformations. Three geometry elements: chemical bonds, pentagons, or hexagons were considered with a maximum of two of these elements being combined at a time. For instance,  the two "5/6 3+3" bonds may be constructed taking into consideration the bridging of one pentagon to one hexagon while the "double 66/66 4+4" takes into consideration the bridging of two hexagons from each molecule. For simplicity, we only considered systems with an even number of intermolecular bonds in this procedure. 
 
We present the HOMO-LUMO gaps at the B3LYP-6-31G(d,p) level as this Hamiltonian is expected to yield closer values to experiment \cite{c60gap}. Note that all calculations were first run without spin polarization, and, in the cases where structures were found to be unstable under these conditions, they were then run considering triplet or quintuplet ground states. This procedure yielded 12 dimers with spin-polarized ground states. In all cases, no initial symmetry constraints were used in the geometry optimizations. 
%, so that the symmetry reported is the e one obtained from the optimized molecular structures. 

 \section*{Results and Discussion}

 The dimer structures found are comprehensively summarized in table \ref{at1} by increasing energy with their symmetry, number of intermolecular bonds, HOMO-LUMO gap, bonding energy, the distance between $\mathrm{C}_{60}$ molecules and a label indicating the minima origin. A graphical rendering for all the bonding schemes found on these dimers together with some others that were found by AMK but are not stable at the DFT level, can be found on the SI, table S3. Relevantly, of the 12 dimers previously reported just the 6/66 2+4 cycloaddition reported by Matsuzawa et al. \cite{Matsuzawa_dimeros} was found not to be stable in our study, although it was found by AMK (AMK minimum 3097) indicating that at the semiempirical PM7 level, this dimer is stable. 
 
Only one covalent dimer with a bonding energy of $-0.25$ eV was found to be thermodynamically more stable, i.e. exothermic, than the isolated $\mathrm{C}_{60}$ molecules. In comparison, the less stable dimer is considerably more endothermic with a  bonding energy of 7.95 eV. As a reference, we also computed the energy of the non-covalent dimer, $-0.36$ eV, which is more stable than the isolated molecule and any covalent dimer. These results match experimental observations since the only, experimentally observed, covalent dimer is fruit of the 66/66 2+2 cycloaddition \cite{i18,i57,strout1993dim} and at ambient conditions, $\mathrm{C}_{60}$ molecules are not covalently bound in the solid state. More exotic bonding configurations, for instance, corresponding to the 56/56 2+2 cycloaddition \cite{laran2018}, can be found in three-dimensional crystals formed under high-pressure treatments, where the 66/66 2+2 cycloaddition bonding scheme alone cannot be used to bond all molecules in a three-dimensional extended system.

  \setlength{\LTcapwidth}{\textwidth}
\onecolumngrid
 \begin{longtable}{c | c | c |c | c | c | c | c}
 	Minima &	Dimer & Sym. & interm. &  HOMO-LUMO &  $\Delta \mathrm{E}_{Bond} $  & Mol.  & bond \\
 	order	& name	& & bonds &  gap (eV)& (eV/dimer) & dist. \AA & origin \\
 	\hline
	&	$\mathrm{C}_{60}$ & $\mathrm{I}_{\mathrm{h}}$           & 0 & 2.8523   &        -    &  - & - \\
           &   Non covalent &  $\mathrm{C}_{\mathrm{1}}$        & 0 & 2.59561  &   -0.3604    & 9.88 & - \\
 	Min1   &	66/66 2+2  & $\mathrm{D}_{2\mathrm{h}}$         & 2 & 2.50562  &   -0.2494   & 9.05 &  AMK66/L \cite{strout1993dim,adamsprb94_dimeros,i26,osawa_dimeros,Matsuzawa_dimeros} \\  
 	Min2   &	56/66 2+2  & $\mathrm{C}_{\mathrm{s}}$          & 2 & 1.81935  &    0.4989   & 9.07 &  AMK203/L \cite{strout1993dim} \\
 	Min3   &	SB  & $\mathrm{C}_{2\mathrm{h}}$                & 1 & 1.9126*  &    0.7440   & 9.23 &  L \cite{i26}\\
 	Min4   &	 56/65 2+2  & $\mathrm{C}_{2\mathrm{h}}$        & 2 & 1.69636  &    1.2501   & 9.09 &  AMK1473/L \cite{strout1993dim} \\
 	Min5   &	56/56 2+2  & $\mathrm{C}_{2\mathrm{v}}$         & 2 & 1.74996  &    1.2771   & 9.10 & AMK1818/L \cite{strout1993dim} \\   
 	Min6   &	double 66/66 2+2  & $\mathrm{C}_{2\mathrm{v}}$  & 4 & 2.49311  &    1.4683   & 8.64 &  G \\
 	Min7   &	double 56/66 2+2 M2 & $\mathrm{C}_{\mathrm{s}}$ & 4 & 2.45719  &    1.8367   & 8.65 & AMK1269  \\
 	Min8   &	 5/66 3+2 & $\mathrm{C}_{\mathrm{s}}$           & 2 & 1.8779*  &    2.1629   & 8.91 & AMK3333  \\                                                  %%    <->  n_min 8    old 9
 	Min9   & double 56/56 2+2  & $\mathrm{C}_{2\mathrm{v}}$     & 4 & 2.34045  &    2.2527   & 8.66 & G  \\                                                        %%    <->  n_min 9    old 8
 	Min10   & AMK5678  & $\mathrm{C}_{\mathrm{1}}$              & 3 & 1.5557*  &    2.3363   & 8.85 & AMK5678  \\                                                  %%   
 	Min11   & 6/66 3+2  &  $\mathrm{C}_{\mathrm{1}}$            & 2 & 0.74614  &    2.3867   & 8.87 & AMK3615  \\                                                  %%    <->  n_min 11   old 13
 	Min12   & 6/6 4+4 M1 & $\mathrm{C}_{2\mathrm{v}}$           & 2 & 1.97527  &    2.6221   & 8.81 & AMK2506/L \cite{osawa_dimeros,Matsuzawa_dimeros}  \\         %%    <->  n_min 12   old 14
 	Min13   & triple 66/65 2+2  & $\mathrm{D}_{3\mathrm{d}}$    & 6 & 2.39487  &    2.6419   & 8.44 & L \cite{adamsprb94_dimeros,LIU2007_dimeros}  \\              %%    <->  n_min 13   old 11
 	Min14   & 6/6 4+4 M2 & $\mathrm{C}_{\mathrm{1}}$            & 2 & 1.79731  &    2.6441   & 8.81 & AMK2507/L \cite{osawa_dimeros,Matsuzawa_dimeros}  \\         %%    <->  n_min 14   old 15
    Min15   & triple 66/66 2+2  & $\mathrm{D}_{3\mathrm{h}}$    & 6 & 2.4327   &    2.6783   & 8.44 & L \cite{adamsprb94_dimeros,LIU2007_dimeros}  \\              %%    <->  n_min 15   old 12
 	Min16   & 6/6 3+4 M2 &  $\mathrm{C}_{\mathrm{1}}$           & 2 & 1.8916*  &    2.8918   & 8.80 &  AMK4004 \\                                                  %%   
 	Min17   & 6/6 3+3 M3  & $\mathrm{C}_{\mathrm{2}}$           & 2 & 1.8632*  &    3.1388   & 8.77 & AMK5325   \\                                                 %%   
 	Min18 & double 56/66 2+2  M1 & $\mathrm{C}_{\mathrm{s}}$    & 4 & 1.56792  &    3.2546   & 8.69 & AMK2605  \\                                                  %%   
 	Min19   & 5/6 3+4 M2  & $\mathrm{C}_{\mathrm{s}}$           & 2 & 1.01063  &    3.3104   & 8.87 & AMK4307  \\                                                  %%    <->  n_min 19   old 21
 	Min20   & 5/6 3+4 M1  & $\mathrm{C}_{\mathrm{s}}$           & 2 & 1.01798  &    3.3326   & 8.87 &  AMK4321 \\                                                  %%    <->  n_min 20   old 22
 	Min21   & 6/56 3+2 M2 &  $\mathrm{C}_{\mathrm{1}}$          & 2 & 1.6369*  &    3.3799   & 8.87 & AMK4646  \\                                                  %%    <->  n_min 21   old 19
 	Min22   & 6/56 3+2 M1 & $\mathrm{C}_{\mathrm{1}}$           & 2 & 1.6207*  &    3.4201   & 8.87 &  AMK4567 \\                                                  %%    <->  n_min 22   old 20
 	Min23   & double 56/65 2+2  & $\mathrm{C}_{\mathrm{s}}$     & 4 & 1.52057  &    3.6534   & 8.71 & AMK3018  \\                                                  %%    
 	Min24   & 5/5 3+3 M1  & $\mathrm{C}_{2\mathrm{h}}$          & 2 & 1.6465*  &    3.6760   & 8.89 &  G \\                                                        %%    
 	Min25   & 6/6 3+4 M1  & $\mathrm{C}_{\mathrm{1}}$           & 2 & 1.3896*  &    3.9138   & 8.77 &  AMK3893 \\                                                  %%    
 	Min26   & double 56/56 2+2 & $\mathrm{C}_{2\mathrm{v}}$     & 4 & 1.22506  &    4.4948   & 8.75 & G   \\                                                       %%    
 	Min27   & AMK6163  & $\mathrm{C}_{1}$                       & 3 & 1.8342*  &    4.6261   & 8.41 &  AMK6163 \\                                                  %%    
 	Min28 &double 56/56 2+2 plus SB&$\mathrm{D}_{5\mathrm{h}}$  & 5 & 0.74260  &    4.8285   & 8.68 &  L \cite{adamsprb94_dimeros} \\                              %%    <->  n_min 28   old 29
 	Min29   & double 6/56 4+2 & $\mathrm{C}_{2\mathrm{h}}$      & 4 & 2.34100  &    5.0253   & 8.62 &  L \cite{strout1993dim}  \\                                  %%    <->  n_min 29   old 28
 	Min30 & 5/56 3+2 plus 6/56 4+2 & $\mathrm{C}_{\mathrm{s}}$  & 4 & 1.52955  &    5.1168   & 8.67 & G \\                                                         %%    
 	Min31   & double 5/5 2+3 &  $\mathrm{C}_{2\mathrm{h}}$      & 4 & 1.10995  &    5.2631   & 8.72 &  G \\                                                        %%    
 	Min32   & 5/6 3+3 plus 56/6 2+3 & $\mathrm{C}_{1}$          & 4 & 1.4211*  &    5.5353   & 8.62 &  AMK5775 \\                                                  %%    
 	Min33   & AMK7077 & $\mathrm{C}_{\mathrm{s}}$               & 4 & 0.87974  &    6.3159   & 8.29 & AMK7077 \\                                                   %%    <->  n_min 33   old 34
 	Min34   & AMK6511 & $\mathrm{C}_{1}$                        & 4 & 1.68194  &    6.4343   & 8.28 &  AMK6511 \\                                                  %%    <->  n_min 34   old 33 
 	Min35   & AMK7610 & $\mathrm{C}_{\mathrm{s}}$               & 4 & 0.95349  &    6.7170   & 8.25 & AMK7610  \\                                                  %%    
 	Min36 & 5/5 3+3 plus 6/6 4+4  & $\mathrm{C}_{2\mathrm{v}}$  & 4 & 0.76546  &    6.8452   & 8.05 & G  \\                                                        %%    <->  n_min 36   old 38
 	Min37    & double 56/65 3+4 & $\mathrm{C}_{2\mathrm{h}}$    & 4 &  0.7584* &    6.9087   & 8.10 & G/AMK\_or \\                                                 %%    <->  n_min 37   old 39
 	Min38 & 5/6 3+4 plus 6/6 4+4  & $\mathrm{C}_{\mathrm{s}}$   & 4 & 0.94070  &    6.9973   & 8.04 & G  \\                                                        %%    <->  n_min 38   old 37
 	Min39   & double 66/66 4+4 &  $\mathrm{D}_{2\mathrm{h}}$    & 4 & 1.64030  &    7.0261   & 8.05 & L  \\                                                        %%    <->  n_min 39   old 36
 	Min40   &	AMK8161  & $\mathrm{C}_{\mathrm{s}}$            & 4 & 2.51216  &    7.5823   & 8.28 & AMK8161  \\                                                  %%    <->  n_min 40   old 41
 	Min41  & double 66/66 4+4  &  $\mathrm{D}_{2\mathrm{h}}$    & 6 & 2.01174  &    7.9499   & 8.16 &  G \\                                                        %%    <->  n_min 41   old 40
 	&	plus  66/66 2+2  &   &  &   &   & &  \\                                                    \caption{Energetics and distances of $\mathrm{C}_{60}$+$\mathrm{C}_{60}$ dimers computed at TPSS-Def2-TZVPP-D3BJ/B3LYP-6-31G(d,p)-D3BJ level. The HOMO-LUMO gaps presented were computed with B3LYP-6-31G(d,p)-D3BJ with * indicating that the presented value is an average of the $\alpha$ and $\beta$ spin channels gaps (table S2 in SI has the used values). Letter L indicates a structure found in Literature while AMK and G indicate the structure was found using, respectively, AutoMeKin or geometric considerations. The number that follows AMK is an internal reference produced by AMK during the generation.}
 	\label{at1}
 \end{longtable}   

     \begin{figure}[H]
	\centering
	\includegraphics[scale=0.3]{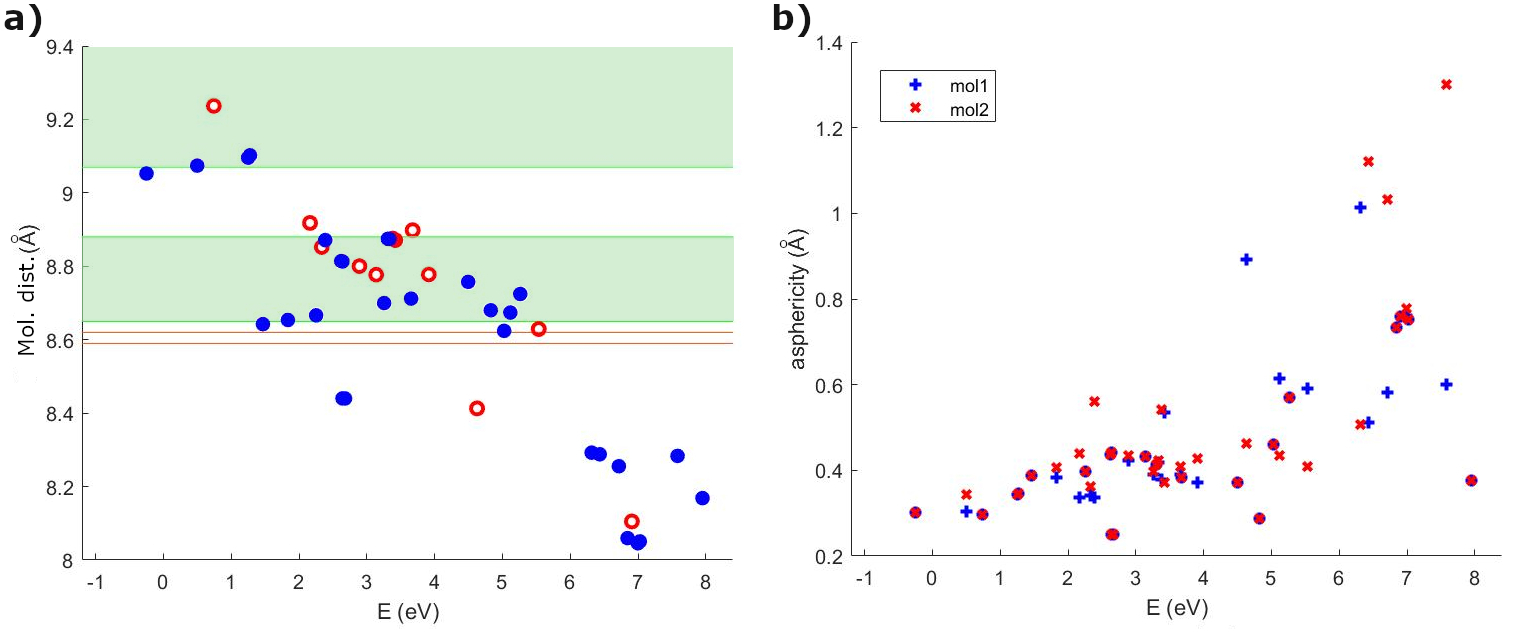}
	\caption{a) Intermolecular  distances as a function of the bonding energy computed at TPSS-Def2-TZVPP-D3BJ/B3LYP-6-31G(d,p)-D3BJ level of theory. Red circles indicate open-shell ground states and blue full circles closed-shell ground states. Red lines and green regions correspond to the experimental molecular distances reported by Yamanaka et al.\cite{i12} and Buga et al.\cite{BUGA2005} respectively. b) Asphericity as a function of the bonding energy. Blue cruxes and red x represent the two different $\mathrm{C}_{60}$ molecules in each dimer.}
 \label{fig1}
\end{figure}

\twocolumngrid
Relevantly, with more extreme pressures, higher energy minima can be reached with reduced $\mathrm{C}_{60}$ distances, fitting the trend shown in figure \ref{fig1} a) where the distances between molecules show a monotonous decrease with increasing bonding energies. Furthermore, several of our dimers have shorter distances than reported coalescence products, see for instance the report by Strout et al. \cite{strout1993dim}. This indicates, that very short distances, below 8.3 \AA, may alternatively be explained by $\mathrm{C}_{60}$ molecules bonded by the patterns reported here. Also, it is important to notice that although all the bonding schemes yield lower HOMO-LUMO gaps compared to $\mathrm{C}_{60}$, no clear trend between this quantity and energy, number of bonds or intermolecular distance was found. 

Interestingly, some of the found dimers are minima only with spin-polarized, triplet or quintuplet, wave-functions. If the minimization is performed without spin-polarization, the dimers either break into $\mathrm{C}_{60}$ molecules or transform into the more stable 66/66 2+2 cycloaddition dimer. Dimers with spin-polarized ground states, in table \ref{at1}, are marked with a star (*) in the HOMO-LUMO value. In addition, the value reported corresponds to the average gap for $\alpha$ and $\beta$ electrons, all values are available on the SI, table S2. Minima 3, 8, 10, 27 and 37 have triplet ground states while minima 16, 17, 21, 22, 24, 25 and 32 have quintuplet ground states. This sheds light on the potential importance of spin-polarization in high-pressure polymerized $\mathrm{C}_{60}$ since some of these bonds may be present in high-pressure formed structures and thus, models built with them probably require spin-polarized Hamiltonians to be properly accounted for. In fact, Bernasconi et al. \cite{bernasconi} took this into consideration while computing three-dimensional $\mathrm{C}_{60}$ polymers noticing that some of the modeled structures had more stable spin-polarized ground states.
 
  After the dimer formation, all the $\mathrm{C}_{60}$ molecules present some degree of deformation with respect to their original structure. To quantify the degree of molecular deformation we computed the asphericity of each molecule defined as the largest difference between the distance to the molecular center of mass for each atom and the same distance obtained for the pristine $\mathrm{C}_{60}$, 3.51 \AA. Asphericity values are summarized in table \ref{at1b}, relevantly, the dimers with higher symmetries than $\mathrm{C}_{\mathrm{s}}$ have the same degree of deformation in both molecules, i.e. the same asphericity. In this case, asphericity values range from 0.25 \AA \space for minima 13 and 15 to 0.75 \AA \space for minima 37 and 39. Figure \ref{fig1} b) shows the asphericity values as a function of the bonding energy. The asphericity qualitatively increases with increasingly endothermic bonding energies, indicating how higher deformation is clearly related to higher energetics and to the more extreme conditions needed for the synthesis of structures displaying these bonding schemes. In fact, the largest asphericity is found for one of the most energetic dimers, minimum 40, with a value of 1.30 \AA.

 	 \begin{table}
\begin{tabular}{c | c | c }
 	Minima &	asphericity & asphericity  \\
    &	of molecule 1  (\AA)& of molecule 2 (\AA)  \\
 	\hline
 	Min1    &    0.3018     &  0.3018 \\
 	Min2    &    0.3034     &  0.3431 \\
 	Min3    &    0.2977     &  0.2977 \\
 	Min4    &    0.3439     &  0.3439 \\
 	Min5    &    0.3470     &  0.3470 \\
 	Min6    &    0.3889     &  0.3889 \\
 	Min7    &    0.3831     &  0.4056 \\
 	Min8    &	 0.3376     &  0.4389 \\  %%    <->  n_min 8    old 9	
 	Min9    &    0.3980     &  0.3980 \\  %%	<->  n_min 9    old 8
 	Min10   &    0.3406     & 0.3617  \\  %%   
 	Min11   &   0.3365      & 0.5613  \\  %%    <->  n_min 11   old 13
 	Min12   &   0.4364      & 0.4364  \\  %%    <->  n_min 12   old 14
 	Min13   &   0.2504      &  0.2504 \\  %%    <->  n_min 13   old 11
 	Min14   &   0.4421      & 0.4421  \\  %%    <->  n_min 14   old 15
        Min15   &   0.2513      &  0.2513 \\  %%    <->  n_min 15   old 12
 	Min16   &   0.4239      & 0.4353  \\  %%   
 	Min17   &   0.4334      & 0.4334  \\  %%   
 	Min18   &   0.3909      & 0.3971  \\  %%   
 	Min19   &   0.4140      & 0.4147  \\  %%    <->  n_min 19   old 21
 	Min20   &   0.4191      &  0.4241 \\  %%    <->  n_min 20   old 22
 	Min21   &   0.3786      & 0.5421  \\  %%    <->  n_min 21   old 19
 	Min22   &   0.5351      & 0.3725  \\  %%    <->  n_min 22   old 20
 	Min23   &   0.3899      &  0.4091 \\  %%    
 	Min24   &   0.3826      &  0.3826 \\  %%    
 	Min25   &   0.3717      &  0.4273 \\  %%    
 	Min26   &   0.3726      &  0.3726 \\  %%    
 	Min27   &   0.8917      &  0.4637 \\  %%    
 	Min28   &    0.2869     &  0.2869 \\  %%    <->  n_min 28   old 29
 	Min29   &   0.4599      &  0.4599 \\  %%    <->  n_min 29   old 28
 	Min30   &    0.6149     &  0.4343 \\  %%    
 	Min31   &    0.5692     &  0.5692 \\  %%    
 	Min32   &    0.5916     &  0.4099 \\  %%    
 	Min33   &    1.0139     &  0.5077 \\  %%    <->  n_min 33   old 34
 	Min34   &    0.5109     &  1.1206 \\  %%    <->  n_min 34   old 33 
 	Min35   &    0.5816     &  1.0315 \\  %%    
 	Min36   &    0.7341     &  0.7341 \\  %%    <->  n_min 36   old 38
 	Min37   &    0.7595     &  0.7595 \\  %%    <->  n_min 37   old 39
 	Min38   &    0.7623     &  0.7786 \\  %%    <->  n_min 38   old 37
 	Min39   &    0.7520     &  0.7520 \\  %%    <->  n_min 39   old 36
 	Min40   &    0.6009     &  1.3002 \\  %%    <->  n_min 40   old 41
 	Min41   &    0.3764     &  0.3764 \\  %%    <->  n_min 41   old 40
                                          
 		\end{tabular}
 	\caption{Asphericities of each molecule in the $\mathrm{C}_{60}$+$\mathrm{C}_{60}$ dimers.}
 	\label{at1b}
 \end{table}

\begin{figure}[t!]
  	%	\hspace*{-1.5cm} %shiftar à esquerda
  	\centering
  	\includegraphics[scale=1.2]{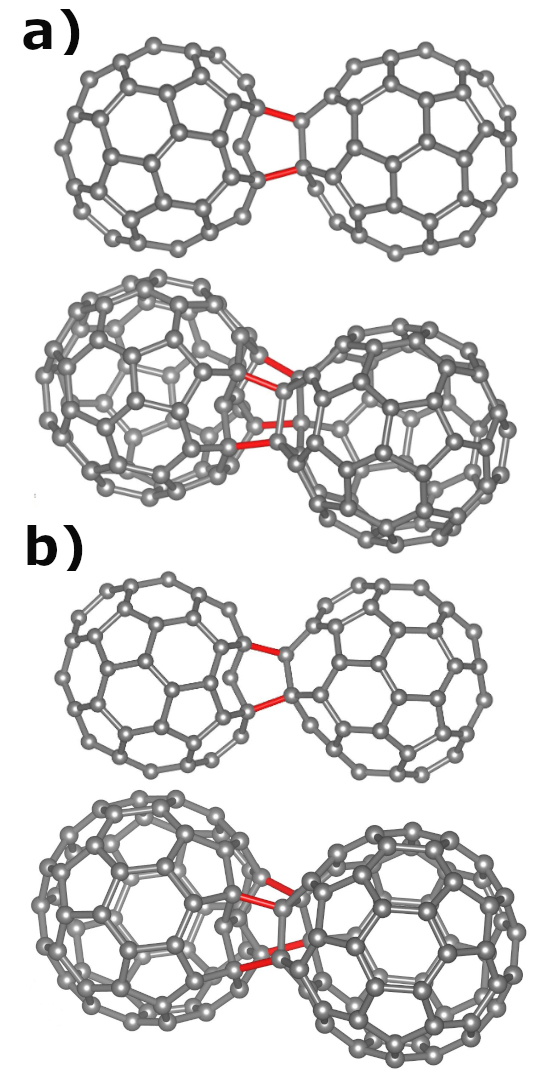}
  	\caption{Two distinct views of selected exotic bonding dimers found by AMK. a) Dimer AMK7077, min 33. b) Dimer AMK7610, min35.}
  	\label{fig2}
  \end{figure}

  AMK was crucial in finding exotic bonding schemes, like Minima 33 and 35 where one of the molecules is pinched allowing the bonding between one pentagon (for minimum 33 or hexagon for minimum 35) of one molecule to one hexagon and one pentagon of the other molecule, see figures \ref{fig2} a) and b). This pinching implies a high asphericity value in the pinched molecule, $\sim$1.0 \AA, while the other molecule has nearly half that value in both dimers, see table \ref{at1b}. In both cases, this translates to remarkably short distances, $\sim$8.25 \AA, and is illustrative for dimers with short bonding distances, $\leq 8.4$ \AA. Furthermore, AMK found another peculiar bond, the double 5/5 2+3 cycloaddition (minimum 31), that was already present in a novel clathrate structure with very similar distances to the ones presented here \cite{LARANJEIRA2022}.
  %between 8.0 and 8.4 \AA.

%***The graphulerene molecular bonding distances for the single bond and 56/65 2+2 cycloaddition are 9.23 and 9.09 \AA, respectively \%cite{naturemgc60_22,naturemgc60_23}, which are in striking agreement with our result further proving the validity of our approach.***

 Buga et al. \cite{BUGA2005} presented a series of six experimental crystalline phases obtained by pressurizing $\mathrm{C}_{60}$ between 11.5 and 13 GPa. In their report, structures are body-centered orthorhombic with the Immm space group ($\Gamma^v_oD^{25}_{2h}$ in Sch\"{o}nflies notation) and presented their lattice parameters although no actual structure was proposed. Since their structures are body-centered orthorhombic, each of them has three distinct $\mathrm{C}_{60}$ distances, table \ref{at2}. In figure \ref{fig1} a), we show a green area where these intermolecular distances lie. The longest distance found in this study is $\sim$9.2 \AA, in minimum 3. Note that as distances by different computational methods may differ, if slightly, we have considered a $\pm0.2$ \AA \space of length "flexibility" to our intermolecular distance values. In addition, molecules may also have additional deformations while in the crystalline phase, therefore hindering the direct correspondence between dimer distances and crystalline distances. This is observed, for instance, in the 66/66 2+2 cycloaddition dimer that shows different experimental distances lying between 9.02 and 9.18 \AA \space \cite{i17,i6}. Hence, experimental intermolecular distances larger than 9.3 \AA \space in table \ref{at2}, may be due to either a single bond or no bond between neighboring molecules. Interestingly, several experimental distances between 8.65 and 8.80 \AA \space in table \ref{at2} are common distances in our dimers. Considering this, it seems likely that some of our bonding schemes are responsible for the distances found in these experimental phases. 

 \begin{table}
 		\begin{tabular}{c | c | c |c | c | c  | c } 
 			Sample Nº& 1 & 2 & 3 & 4 & 5 & 6 \\
 			\hline
 			distance 1 (\AA)	& 9.07 & 8.73 & 8.69 & 8.68 & 8.67 & 8.65 \\
 			distance 2 (\AA)	& 9.47 & 9.16 & 8.88 & 8.84 & 8.81 & 8.78 \\
 			distance 3 (\AA)	& 9.19 & 9.05 & 8.88 & 8.85 & 8.83 & 8.79 \\	
 		\end{tabular} 
 	\caption{Intermolecular distances from the structures reported in \cite{BUGA2005}.}
 	\label{at2}
 \end{table}

In addition, there are other structures in the literature still to be resolved, like the one reported by Yamanaka et al. \cite{i12}. Although a complete crystalline structure was proposed, the structure changed considerably when optimized with DFT without constraints \cite{i13,bernasconi}. Namely, Yamanaka et al. derived experimental lattice parameters of 7.86, 8.59 and 12.73 \AA \space and a body-centered orthorhombic structure, with corresponding distances of 7.86, 8.59 and 8.62 \AA, indicated in figure \ref{fig1} a) by the red lines. The authors assigned these distances to double 66/66 4+4, no bond and 6/6 3+3 cycloaddition, respectively. Indeed, the 6/6 3+3 cycloaddition dimer, Min 17, and 66/66 4+4 cycloaddition dimer, min 39, have bonding distances in the appropriate range to explain this experimental data. Nevertheless, these bonding schemes were already used by Zipoli and Bernasconi \cite{bernasconi} to construct crystalline structures and fail to reproduce the experimental lattice cell. Thus, other bonding schemes must be considered. We present here several potential alternatives within the observed range of distances. 
%, hence having the possibility to elucidate the experimental observations.  

Conversely, although the comparison of the distance is a simple, yet effective tool, we would like to emphasize that bonding patterns, which are stable in dimers may not be stable or not even geometrically possible in specific crystalline structures. For instance, minimum 4, although geometrically capable to bond to all the neighboring molecules in a crystalline fcc structure, turns out to be unstable in such configuration \cite{laran2017,laran2018}. On the other hand, the same minimum is present in other reported structures, such as in a 3D-rhombohedral phase \cite{i68} and in the recent graphullerene \cite{naturemgc60_23}, proving that the molecular environment as a whole has to be considered to evaluate the existence of these patterns. 

%redundant
%, similar to our unsuccessfully attempt to bond all the nearest neighbors in the crystalline fcc structure \cite{laran2018} .***

\section*{Conclusion}
A comprehensive exploration and in-depth characterization of a large number of $\mathrm{C}_{60}$ dimers built from high energy molecular dynamics trajectories through the AMK methodology was performed. Several new $\mathrm{C}_{60}$ bonding schemes potentially relevant to understand polymeric phases were found and were energetically, structurally and electronically classified, increasing the number of known $\mathrm{C}_{60}$ dimers from 12 to 41. Some of the new dimers have bonding schemes matching experimental intermolecular distances observed on high-pressure high-temperature polymerized $\mathrm{C}_{60}$ yet to be understood, for instance in the works of Buga et al. or Yamanaka et al. \cite{BUGA2005,i12}. Relevantly, 12 of these dimers, i.e. 29 \%, have spin-polarized ground states, which emphasizes that spin-polarization should be explicitly taken into consideration to properly model crystalline structures, especially with metallic or electronic exotic properties. This work paves the way for the exploration of novel extended 1D, 2D and 3D structures based on the bonding schemes presented here obtainable by HPHT treatment or more recent alternative routes \cite{naturemgc60_22, naturemgc60_23}. 
\newline

\section*{Author Contributions}

Conceptualization: M. Melle-Franco and J. Laranjeira; methodology: M. Melle-Franco, J. Laranjeira and E. Martínez-Núñez; software: M. Melle-Franco and E. Martínez-Núñez; validation: L. Marques, K. Struty\`{n}ski and E. Martínez-Núñez; formal analysis: J. Laranjeira; investigation: J. Laranjeira; resources: M. Melle-Franco, E. Martínez-Núñez and J. Laranjeira; data curation: J. Laranjeira; writing-original draft preparation: J. Laranjeira; writing-review and editing: J. Laranjeira, M. Melle-Franco; visualization: J. Laranjeira; supervision: M. Melle-Franco; funding acquisition: M. Melle-Franco, E. Martínez-Núñez and J. Laranjeira. All authors have read and agreed to the published version of the manuscript.
\begin{acknowledgements}
This work was developed within the scope of the project CICECO-Aveiro Institute of Materials, UIDB/50011/2020, UIDP/50011/2020 \& LA/P/0006/2020, financed by national funds through the FCT/MCTES (PIDDAC) and IF/00894/2015 finances by FCT. J. Laranjeira acknowledges a PhD grant from FCT (SFRH/BD/139327/2018).

This work was partially supported by the Consellería de Cultura, Educaci\'on e Ordenaci\'on Universitaria (Grupo de referencia competitiva ED431C 2021/40), and the Ministerio de Ciencia e Innovaci\'on through Grant \#PID2019-107307RB-I00.

The work has been performed under the project HPC-EUROPA3 (INFRAIA-2016-1-730897), with the support of the EC Research Innovation Action under the H2020 Programme; in particular, the author gratefully acknowledges the support of the computer resources and technical support provided by BSC.

\end{acknowledgements}

\bibliography{referencias.bib}% Produces the bibliography via BibTeX.

\end{document}

\preprint{APS/123-QED}
\onecolumngrid
\section*{Supporting Information}
\section*{A. Bonding notation}
\label{notacao}
The naming of $\mathrm{C}_{60}$+$\mathrm{C}_{60}$ bonding schemes is done following the so-called cycloaddition rules which state how different rings react in organic chemistry in order to fuse. Considering the formation of a new ring of bonds between two molecules the bonding scheme is classified by the geometrical figure (or combination of geometrical figures) that is facing each molecule and by the number of atoms from each molecule that is involved in it. If we consider the more common, and stable, $\mathrm{C}_{60}$ dimer, the intermolecular bonding follows a 66/66 2+2 cycloaddition: 2+2 because from each molecule there are two bonds involved (one intramolecular and one intermolecular), and  66/66 since the two carbon atoms from each molecule are covalently bonded to the other by the two carbon atoms shared by two hexagons  (hence 66 for each molecule). We may also consider that the alignment is done just by one geometrical figure from each molecule. If we consider the alignment between two hexagons, we will have a 6/6 n+m cycloaddition. With n, and m being integers smaller than 4.

\section*{B. Van der Waals correction and exchange-correlation comparison}
 \label{methods}
 \twocolumngrid
 
 \begin{figure}[H]
		\hspace*{-1.5cm} %shiftar à esquerda
	\centering
	\includegraphics[scale=0.35]{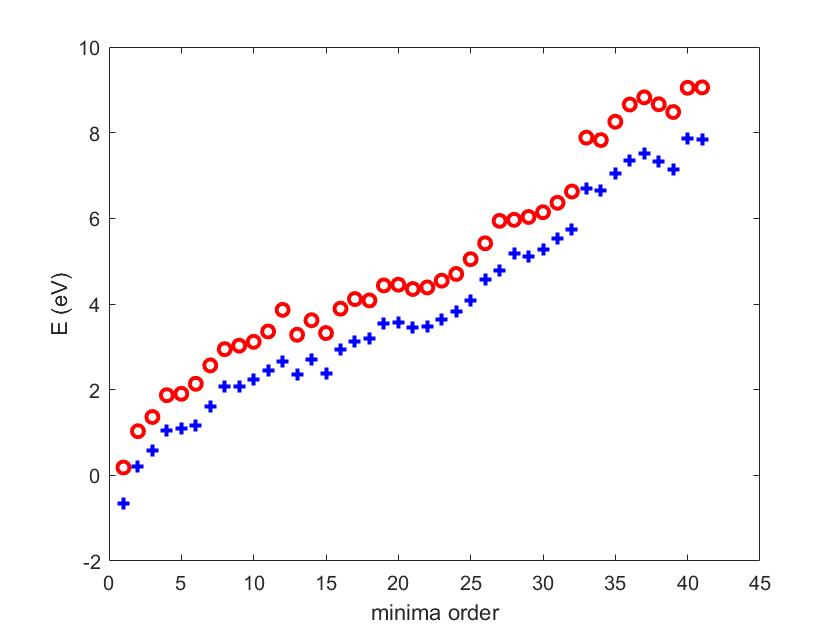}
	\caption{Minima bonding energy in eV per dimer vs minima order computed with B3LYP-6-31G(d,p) Hamiltonian (red circles) and augmented by the D3 van der Waals correction with finite Becke–Johnson damping (blue cruxes) after optimization at B3LYP-6-31G(d,p)-D3BJ level.}
	\label{b3lyp}
\end{figure}

 \begin{figure}[H]
	%	\hspace*{-1.5cm} %shiftar à esquerda
	\centering
	\includegraphics[scale=0.35]{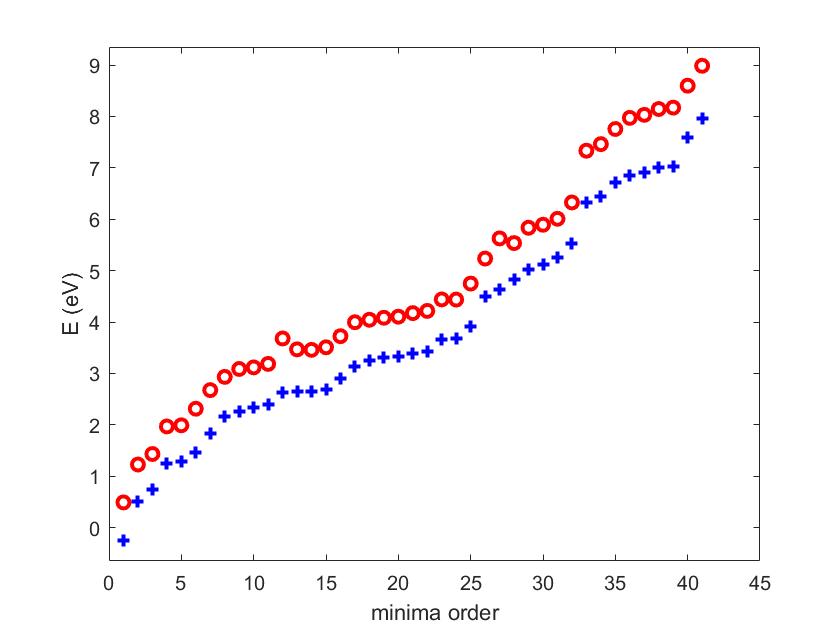}
	\caption{
 Minima bonding energy in eV per dimer vs minima order computed with TPSS-Def2-TZVPP Hamiltonian (red circles) and augmented by the D3 van der Waals correction with finite Becke–Johnson damping (blue cruxes) after optimization at B3LYP-6-31G(d,p)-D3BJ level.}
	\label{tpss}
\end{figure}
\clearpage

\onecolumngrid
 \setlength\LTleft{-2.0cm}
 \setlength{\LTcapwidth}{\textwidth}
 \begin{small} 
 \begin{longtable}{c | c | c | c | c | c | c | c }
 	Minima  & Dimer & Sym. & interm. & $\mathrm{E}_{Bond}^{B3LYP}$ &  $\mathrm{E}_{Bond}^{B3LYP-D3BJ}$ & $\mathrm{E}_{Bond}^{TPSS-D3BJ} $ & $\mathrm{E}_{Bond}^{TPSS}$  \\
 	order	& name	& & bonds &  (eV/dim) &  (eV/dim) & (eV/dim) & (eV/dim) \\
 	\hline
 	Min1   &	66/66 2+2  & $\mathrm{D}_{2\mathrm{h}}$         & 2 &     0.1823 & -0.6541  &   -0.2494 &    0.4900   \\
   	Min2   &	56/66 2+2  & $\mathrm{C}_{\mathrm{s}}$          & 2 &     1.0300 &  0.2055  &    0.4989 &    1.2269   \\
   	Min3   &	SB  & $\mathrm{C}_{2\mathrm{h}}$                & 1 &     1.3613 &  0.5839  &    0.7440 &    1.4314   \\
   	Min4   &	 56/65 2+2  & $\mathrm{C}_{2\mathrm{h}}$        & 2 &     1.8711 &  1.0571  &    1.2501 &    1.9679   \\
   	Min5   &	56/56 2+2  & $\mathrm{C}_{2\mathrm{v}}$         & 2 &     1.9026 &  1.0934  &    1.2771 &    1.9906   \\
   	Min6   &	double 66/66 2+2  & $\mathrm{C}_{2\mathrm{v}}$  & 4 &     2.1376 &  1.1710  &    1.4683 &    2.3126   \\
   	Min7   &	double 56/66 2+2 M2 & $\mathrm{C}_{\mathrm{s}}$ & 4 &     2.5666 &  1.6022  &    1.8367 &    2.6752   \\
	Min8   &	 5/66 3+2 & $\mathrm{C}_{\mathrm{s}}$           & 2 &     2.9455 &  2.0740  &    2.1629 &    2.9319   \\  %%    <->  n_min 8    old 9
   	Min9   & double 56/56 2+2  & $\mathrm{C}_{2\mathrm{v}}$     & 4 &     3.0279 &  2.0741  &    2.2527 &    3.0859   \\  %%    <->  n_min 9    old 8
   	Min10   & AMK5678  & $\mathrm{C}_{\mathrm{1}}$              & 3 &     3.1171 &  2.2272  &    2.3363 &    3.1170   \\  %%   
   	Min11   & 6/66 3+2  &  $\mathrm{C}_{\mathrm{1}}$            & 2 &     3.3582 &  2.4498  &    2.3867 &    3.1851   \\  %%    <->  n_min 11   old 13
   	Min12   & 6/6 4+4 M1 & $\mathrm{C}_{2\mathrm{v}}$           & 2 &     3.8631 &  2.6621  &    2.6221 &    3.6790   \\  %%    <->  n_min 12   old 14
   	Min13   & triple 66/65 2+2  & $\mathrm{D}_{3\mathrm{d}}$    & 6 &     3.2816 &  2.3517  &    2.6419 &    3.4678   \\  %%    <->  n_min 13   old 11
   	Min14   & 6/6 4+4 M2 & $\mathrm{C}_{\mathrm{1}}$            & 2 &     3.6203 &  2.6950  &    2.6441 &    3.4582   \\  %%    <->  n_min 14   old 15
    Min15   & triple 66/66 2+2  & $\mathrm{D}_{3\mathrm{h}}$    & 6 &     3.3219 &  2.3890  &    2.6783 &    3.5086   \\  %%    <->  n_min 15   old 12
   	Min16   & 6/6 3+4 M2 &  $\mathrm{C}_{\mathrm{1}}$           & 2 &     3.8909 &  2.9370  &    2.8918 &    3.7243   \\  %%   
   	Min17   & 6/6 3+3 M3  & $\mathrm{C}_{\mathrm{2}}$           & 2 &     4.1152 &  3.1303  &    3.1388 &    3.9958   \\  %%   
   	Min18 & double 56/66 2+2  M1 & $\mathrm{C}_{\mathrm{s}}$    & 4 &     4.0830 &  3.1872  &    3.2546 &    4.0437   \\  %%   
   	Min19   & 5/6 3+4 M2  & $\mathrm{C}_{\mathrm{s}}$           & 2 &     4.4326 &  3.5528  &    3.3104 &    4.0827   \\  %%    <->  n_min 19   old 21
   	Min20   & 5/6 3+4 M1  & $\mathrm{C}_{\mathrm{s}}$           & 2 &     4.4504 &  3.5723  &    3.3326 &    4.1037   \\  %%    <->  n_min 20   old 22
   	Min21   & 6/56 3+2 M2 &  $\mathrm{C}_{\mathrm{1}}$          & 2 &     4.3499 &  3.4470  &    3.3799 &    4.1725   \\  %%    <->  n_min 21   old 19
   	Min22   & 6/56 3+2 M1 & $\mathrm{C}_{\mathrm{1}}$           & 2 &     4.3866 &  3.4796  &    3.4201 &    4.2159   \\  %%    <->  n_min 22   old 20
   	Min23   & double 56/65 2+2  & $\mathrm{C}_{\mathrm{s}}$     & 4 &     4.5444 &  3.6441  &    3.6534 &    4.4392   \\  %%    
   	Min24   & 5/5 3+3 M1  & $\mathrm{C}_{2\mathrm{h}}$          & 2 &     4.7005 &  3.8357  &    3.6760 &    4.4361   \\  %%    
   	Min25   & 6/6 3+4 M1  & $\mathrm{C}_{\mathrm{1}}$           & 2 &     5.0466 &  4.0917  &    3.9138 &    4.7480   \\  %%    
   	Min26   & double 56/56 2+2 & $\mathrm{C}_{2\mathrm{v}}$     & 4 &     5.4181 &  4.5818  &    4.4948 &    5.2343   \\  %%    
   	Min27   & AMK6163  & $\mathrm{C}_{1}$                       & 3 &     5.9410 &  4.7803  &    4.6261 &    5.6268   \\  %%    
   	Min28 &double 56/56 2+2 plus SB&$\mathrm{D}_{5\mathrm{h}}$  & 5 &     5.9654 &  5.1727  &    4.8285 &    5.5361   \\  %%    <->  n_min 28   old 29
   	Min29   & double 6/56 4+2 & $\mathrm{C}_{2\mathrm{h}}$      & 4 &     6.0320 &  5.1107  &    5.0253 &    5.8338   \\  %%    <->  n_min 29   old 28
   	Min30 & 5/56 3+2 plus 6/56 4+2 & $\mathrm{C}_{\mathrm{s}}$  & 4 &     6.1424 &  5.2653  &    5.1168 &    5.8918   \\  %%    
   	Min31   & double 5/5 2+3 &  $\mathrm{C}_{2\mathrm{h}}$      & 4 &     6.3630 &  5.5234  &    5.2631 &    6.0079   \\  %%    
   	Min32   & 5/6 3+3 plus 56/6 2+3 & $\mathrm{C}_{1}$          & 4 &     6.6255 &  5.7290  &    5.5353 &    6.3257   \\  %%    
   	Min33   & AMK7077 & $\mathrm{C}_{\mathrm{s}}$               & 4 &     7.8846 &  6.7070  &    6.3159 &    7.3327   \\  %%    <->  n_min 33   old 34
   	Min34   & AMK6511 & $\mathrm{C}_{1}$                        & 4 &     7.8269 &  6.6398  &    6.4343 &    7.4601   \\  %%    <->  n_min 34   old 33 
   	Min35   & AMK7610 & $\mathrm{C}_{\mathrm{s}}$               & 4 &     8.2559 &  7.0520  &    6.7170 &    7.7550   \\  %%    
	Min36 & 5/5 3+3 plus 6/6 4+4  & $\mathrm{C}_{2\mathrm{v}}$  & 4 &     8.6573 &  7.3448  &    6.8452 &    7.9718   \\  %%    <->  n_min 36   old 38
   	Min37    & double 56/65 3+4 & $\mathrm{C}_{2\mathrm{h}}$    & 4 &     8.8244 &  7.5147  &    6.9087 &    8.0303   \\  %%    <->  n_min 37   old 39
   	Min38 & 5/6 3+4 plus 6/6 4+4  & $\mathrm{C}_{\mathrm{s}}$   & 4 &     8.6623 &  7.3213  &    6.9973 &    8.1441   \\  %%    <->  n_min 38   old 37
   	Min39   & double 66/66 4+4 &  $\mathrm{D}_{2\mathrm{h}}$    & 4 &     8.4821 &  7.1399  &    7.0261 &    8.1709   \\  %%    <->  n_min 39   old 36
   	Min40   &	AMK8161  & $\mathrm{C}_{\mathrm{s}}$            & 4 &     9.0487 &  7.8726  &   7.5823  &    8.5970   \\  %%    <->  n_min 40   old 41	
   	Min41  & double 66/66 4+4  &  $\mathrm{D}_{2\mathrm{h}}$    & 6 &     9.0574 &  7.8481  &    7.9499 &    8.9845   \\  %%    <->  n_min 41   old 40
   	&	plus  66/66 2+2  &  &  &  &   &  & \\                                                                             
  
 	\caption*{Table S1. Bonding energy of $\mathrm{C}_{60}$+$\mathrm{C}_{60}$ dimers optimized with B3LYP-6-31G(d,p) augmented with D3 van der Waals correction with finite Becke–Johnson damping and computed with B3LYP-6-31G(d,p) and TPSS-Def2-TZVPP/B3LYP-6-31G(d,p)-D3BJ, with and without D3 van der Waals correction with finite Becke–Johnson damping augmentation. }
 	\label{t_tpssb3lyp}
 \end{longtable}   
 	\end{small}

\clearpage

\section*{C. Complete HOMO-LUMO gap information}
\label{homolumo}

   \begin{small} 
   \setlength{\LTcapwidth}{\textwidth}
   \begin{longtable}{c | c | c |c | c | c | c }
 	Minima &	Dimer & Sym. & interm.  &  HOMO-LUMO& $\Delta \mathrm{E}_{Bond} $  & Mol.  \\
 	order	& name	& & bonds &    gap (eV) & (eV/dim) & dist. \AA \\
 	\hline
  	&	$\mathrm{C}_{60}$ & $\mathrm{I}_{\mathrm{h}}$           & 0  &  2.8523             &        -    &       \\
   	Min1   &	66/66 2+2  & $\mathrm{D}_{2\mathrm{h}}$         & 2  &  2.50562            &  -0.6541    & 9.05 \\
   	Min2   &	56/66 2+2  & $\mathrm{C}_{\mathrm{s}}$          & 2  &  1.81935            &   0.2055    & 9.07 \\
   	Min3   &	SB  & $\mathrm{C}_{2\mathrm{h}}$                & 1  &  1.84359  1.98155   &   0.5839    & 9.23 \\
   	Min4   &	 56/65 2+2  & $\mathrm{C}_{2\mathrm{h}}$        & 2  &  1.69636            &   1.0571    & 9.09 \\
   	Min5   &	56/56 2+2  & $\mathrm{C}_{2\mathrm{v}}$         & 2  &  1.74996            &   1.0934    & 9.10 \\
   	Min6   &	double 66/66 2+2  & $\mathrm{C}_{2\mathrm{v}}$  & 4  &  2.49311            &   1.1710    & 8.64 \\
   	Min7   &	double 56/66 2+2 M2 & $\mathrm{C}_{\mathrm{s}}$ & 4  &  2.45719            &   1.6022    & 8.65 \\
 	Min8   &	 5/66 3+2 & $\mathrm{C}_{\mathrm{s}}$           & 2  &  1.91651 1.83923    &   2.0740    & 8.91 \\  %%    <->  n_min 8    old 9
   	Min9   & double 56/56 2+2  & $\mathrm{C}_{2\mathrm{v}}$     & 4  &  2.34045            &   2.0741    & 8.66 \\  %%	  <->  n_min 9    old 8
   	Min10   & AMK5678  & $\mathrm{C}_{\mathrm{1}}$              & 3  &  1.41609  1.69528   &   2.2272    & 8.85 \\  %%   
   	Min11   & 6/66 3+2  &  $\mathrm{C}_{\mathrm{1}}$            & 2  &  0.74614            &   2.4498    & 8.87 \\  %%    <->  n_min 11   old 13
   	Min12   & 6/6 4+4 M1 & $\mathrm{C}_{2\mathrm{v}}$           & 2  &  1.97527            &   2.6621    & 8.81 \\  %%    <->  n_min 12   old 14
   	Min13   & triple 66/65 2+2  & $\mathrm{D}_{3\mathrm{d}}$    & 6  &  2.39487            &   2.3517    & 8.44 \\  %%    <->  n_min 13   old 11
   	Min14   & 6/6 4+4 M2 & $\mathrm{C}_{\mathrm{1}}$            & 2  &  1.79731            &   2.6950    & 8.81 \\  %%    <->  n_min 14   old 15
    Min15   & triple 66/66 2+2  & $\mathrm{D}_{3\mathrm{h}}$    & 6  &  2.4327             &   2.3890    & 8.44 \\  %%    <->  n_min 15   old 12
   	Min16   & 6/6 3+4 M2 &  $\mathrm{C}_{\mathrm{1}}$           & 2  &  1.78589 1.99733    &   2.9370    & 8.80 \\  %%   
   	Min17   & 6/6 3+3 M3  & $\mathrm{C}_{\mathrm{2}}$           & 2  &  1.78318 1.94318    &   3.1303    & 8.77 \\  %%   
   	Min18 & double 56/66 2+2  M1 & $\mathrm{C}_{\mathrm{s}}$    & 4  &  1.56792            &   3.1872    & 8.69 \\  %%   
   	Min19   & 5/6 3+4 M2  & $\mathrm{C}_{\mathrm{s}}$           & 2  &  1.01063            &   3.5528    & 8.87 \\  %%    <->  n_min 19   old 21
   	Min20   & 5/6 3+4 M1  & $\mathrm{C}_{\mathrm{s}}$           & 2  &  1.01798            &   3.5723    & 8.87 \\  %%    <->  n_min 20   old 22
   	Min21   & 6/56 3+2 M2 &  $\mathrm{C}_{\mathrm{1}}$          & 2  &  1.53201 1.74182    &   3.4470    & 8.87 \\  %%    <->  n_min 21   old 19
   	Min22   & 6/56 3+2 M1 & $\mathrm{C}_{\mathrm{1}}$           & 2  &  1.51841 1.72304    &   3.4796    & 8.87 \\  %%    <->  n_min 22   old 20
   	Min23   & double 56/65 2+2  & $\mathrm{C}_{\mathrm{s}}$     & 4  &  1.52057            &   3.6441    & 8.71 \\  %%    
   	Min24   & 5/5 3+3 M1  & $\mathrm{C}_{2\mathrm{h}}$          & 2  &  1.8637 1.4294      &   3.8357    & 8.89 \\  %%    
   	Min25   & 6/6 3+4 M1  & $\mathrm{C}_{\mathrm{1}}$           & 2  &  1.28058  1.49854   &   4.0917    & 8.77 \\  %%    
   	Min26   & double 56/56 2+2 & $\mathrm{C}_{2\mathrm{v}}$     & 4  &  1.22506            &   4.5818    & 8.75 \\  %%    
   	Min27   & AMK6163  & $\mathrm{C}_{1}$                       & 3  &  1.75270 1.91569    &   4.7803    & 8.41 \\  %%    
   	Min28 &double 56/56 2+2 plus SB&$\mathrm{D}_{5\mathrm{h}}$  & 5  &  0.74260            &   5.1727    & 8.68 \\  %%    <->  n_min 28   old 29
   	Min29   & double 6/56 4+2 & $\mathrm{C}_{2\mathrm{h}}$      & 4  &  2.34100            &   5.1107    & 8.62 \\  %%    <->  n_min 29   old 28
   	Min30 & 5/56 3+2 plus 6/56 4+2 & $\mathrm{C}_{\mathrm{s}}$  & 4  &  1.52955            &   5.2653    & 8.67 \\  %%    
   	Min31   & double 5/5 2+3 &  $\mathrm{C}_{2\mathrm{h}}$      & 4  &  1.10995            &   5.5234    & 8.72 \\  %%    
   	Min32   & 5/6 3+3 plus 56/6 2+3 & $\mathrm{C}_{1}$          & 4  &  1.39868 1.44358    &   5.7290    & 8.62 \\  %%    
   	Min33   & AMK7077 & $\mathrm{C}_{\mathrm{s}}$               & 4  &  0.87974            &   6.7070    & 8.29 \\  %%    <->  n_min 33   old 34
   	Min34   & AMK6511 & $\mathrm{C}_{1}$                        & 4  &  1.68194            &   6.6398    & 8.28 \\  %%    <->  n_min 34   old 33 
   	Min35   & AMK7610 & $\mathrm{C}_{\mathrm{s}}$               & 4  &  0.95349            &   7.0520    & 8.25 \\  %%    
   	Min36 & 5/5 3+3 plus 6/6 4+4  & $\mathrm{C}_{2\mathrm{v}}$  & 4  &  0.76546            &   7.3448    & 8.05 \\  %%    <->  n_min 36   old 38
   	Min37    & double 56/65 3+4 & $\mathrm{C}_{2\mathrm{h}}$    & 4  & 0.85662 0.66015     &   7.5147    & 8.10 \\  %%    <->  n_min 37   old 39
   	Min38 & 5/6 3+4 plus 6/6 4+4  & $\mathrm{C}_{\mathrm{s}}$   & 4  &  0.94070            &   7.3213    & 8.04 \\  %%    <->  n_min 38   old 37
   	Min39   & double 66/66 4+4 &  $\mathrm{D}_{2\mathrm{h}}$    & 4  &  1.64030            &   7.1399    & 8.05 \\  %%    <->  n_min 39   old 36
   	Min40   &	AMK8161  & $\mathrm{C}_{\mathrm{s}}$            & 4  &   2.51216           &   7.8726    & 8.28 \\  %%    <->  n_min 40   old 41
   	Min41  & double 66/66 4+4  &  $\mathrm{D}_{2\mathrm{h}}$    & 6  &  2.01174            &   7.8481    & 8.16 \\  %%    <->  n_min 41   old 40
   	&	plus  66/66 2+2  &  &  &       &   \\                                                                       
   					
 	\caption*{Table S2. Energetics and molecular distances of $\mathrm{C}_{60}$+$\mathrm{C}_{60}$ dimers computed at B3LYP-6-31G(d,p)-D3BJ level of theory.  Letter L indicate structure found in Literature, AMK and G indicate the structure was found using, respectively, AutoMeKin or geometric considerations. The number that follows AMK is simple the energy order of the minimum given by AMK.}
 	\label{tab_gaps}
 \end{longtable}   
 	\end{small}

\clearpage

\section*{D. Bonding Schemes}
\label{bondingscheme}
 \setlength{\LTcapwidth}{\textwidth}
\setlength\LTleft{-1.0cm}
\begin{longtable}{c |c | c | c } 
\caption*{Table S3. $\mathrm{C}_{60}$+$\mathrm{C}_{60}$ dimer names, number of bonds, bonding schemes and origin of the bond. Letter L indicate structure found in Literature, AMK and G indicate that the structure was found using, respectively, AutoMeKin or geometric considerations. The number that follows AMK is simple the energy order of the minimum given by AMK.}\\

	Isomer & interm. & intermolecular &  bond \\
	& bonds &  bonding scheme & origin \\
	\hline
	$\mathrm{C}_{60}$ &  0 &  -  & - \\
	\hline
	   &  &  &      \\
 Min1	66/66 2+2  &  2 & \includegraphics[scale=0.1]{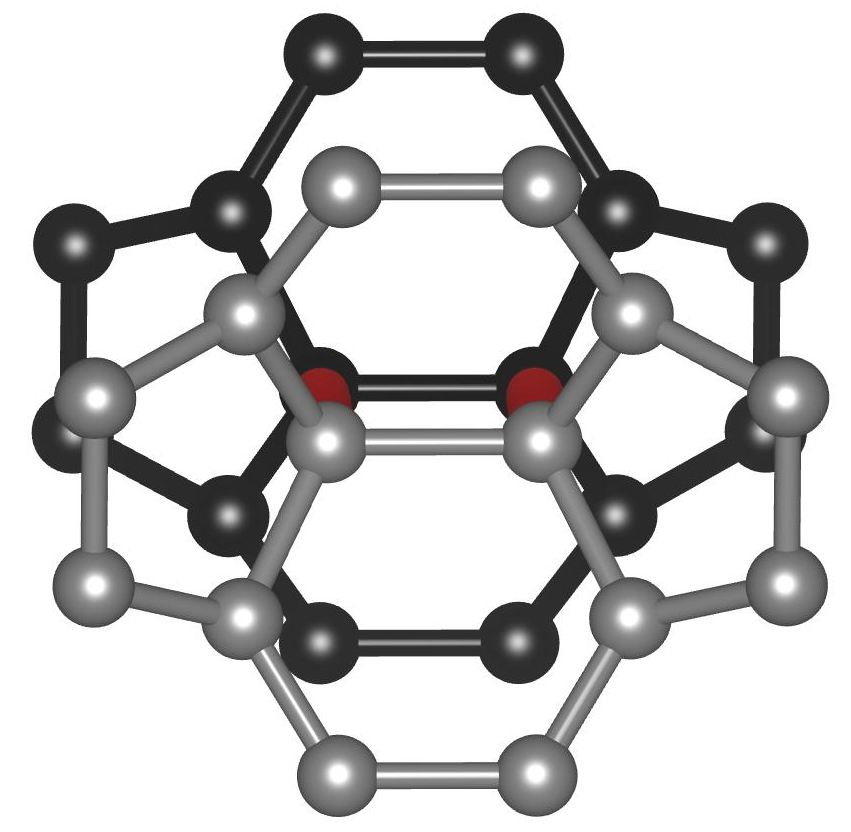}  &     AMK66/L \cite{strout1993dim,adamsprb94_dimeros,i26,osawa_dimeros,Matsuzawa_dimeros} \\
 	   &  &  &      \\
 \hline
 	   &  &  &      \\
 Min2 56/66 2+2  &  2 & \includegraphics[scale=0.1]{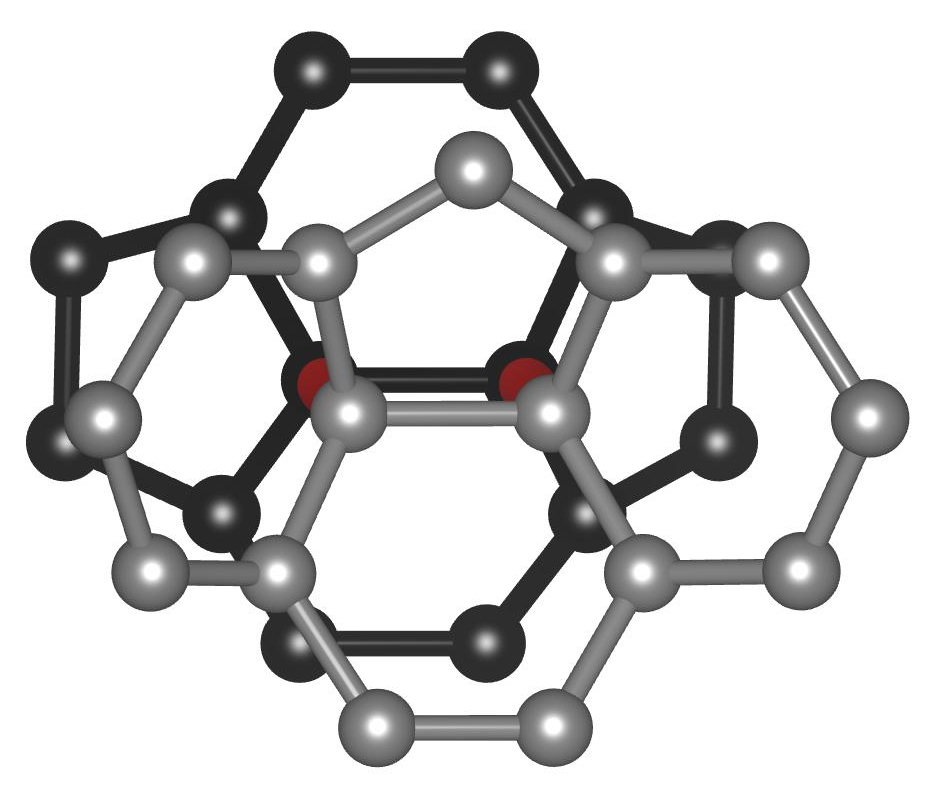} &    AMK203/L \cite{strout1993dim} \\
		   &  &  &      \\
		   \hline
	  %	56/66 2+2 plus 66/66 2+2 & & 4 &  \includegraphics[scale=0.1]{jpgs1/.jpg}  & G \\
		   %	double 56/66 2+2  M3 & & 4 &  &  \includegraphics[scale=0.1]{jpgs1/.jpg}    G \\ 
   &    &  &      \\
    Min3 single bond  &  1 &  \includegraphics[scale=0.1]{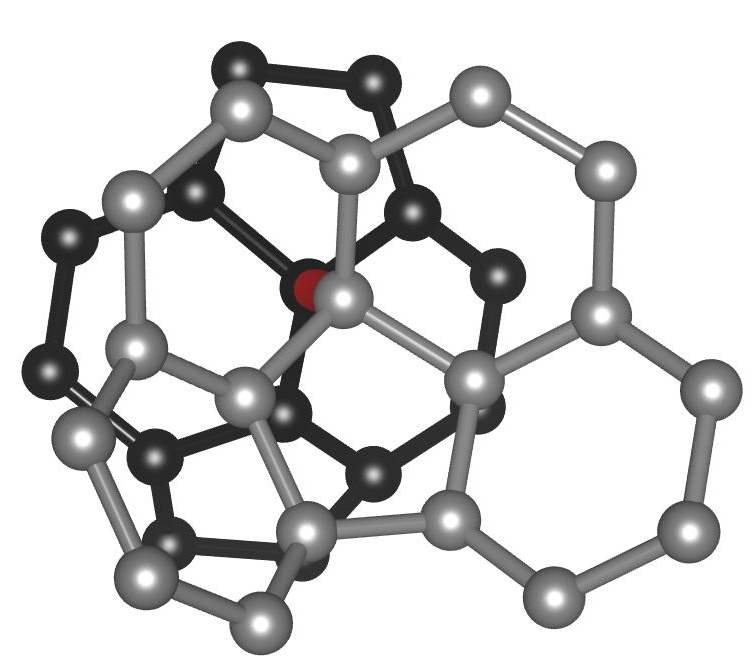}   &  L \cite{i26} \\
		   &  &    &      \\
	\hline
	 &  &  &   \\
	Min4 56/65 2+2  &  2 & \includegraphics[scale=0.1]{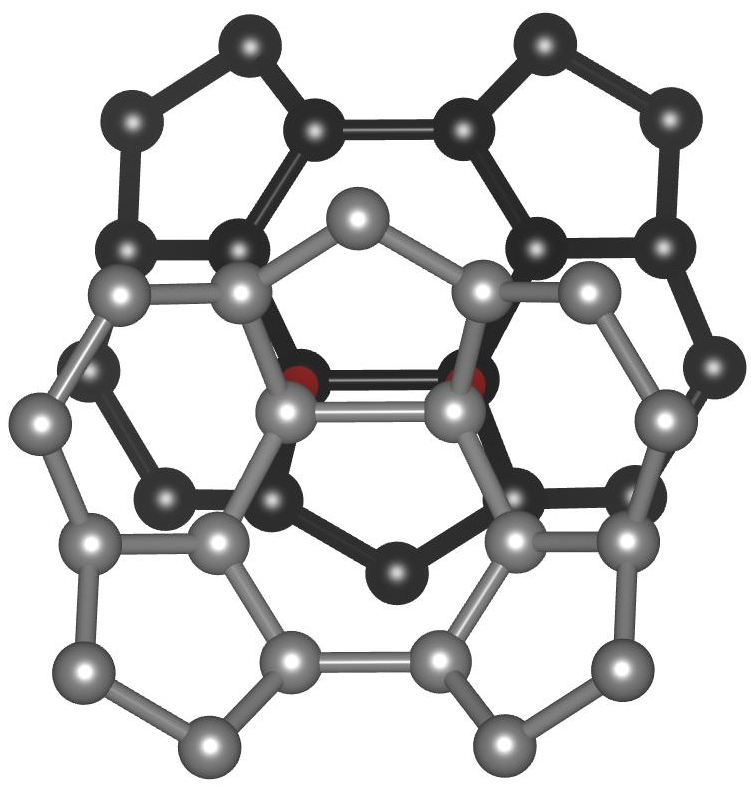} &     AMK1473/L \cite{strout1993dim} \\
	&  &  &   \\
	\hline
	   &  &  &   \\
   Min5	56/56 2+2  & 2 & \includegraphics[scale=0.2]{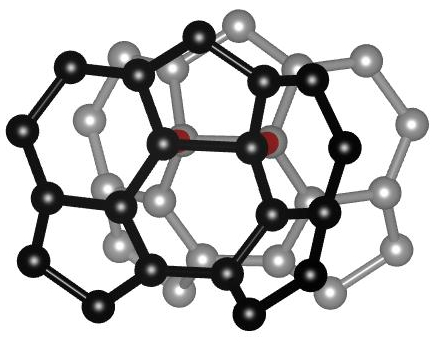}  &    AMK1818/L \cite{strout1993dim}  \\
	   &  &  & \\  
	   \hline
	   &  &    &      \\
	   Min6 double 66/66 2+2 & 4 &  \includegraphics[scale=0.1]{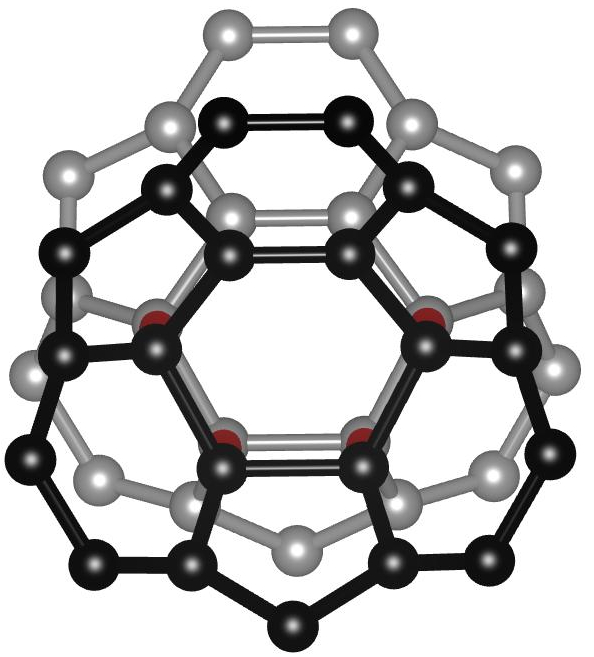}  & G \\
	   \hline
	   &    &  &      \\
	   Min7	double 56/66 2+2  M2  & 4 &  \includegraphics[scale=0.1]{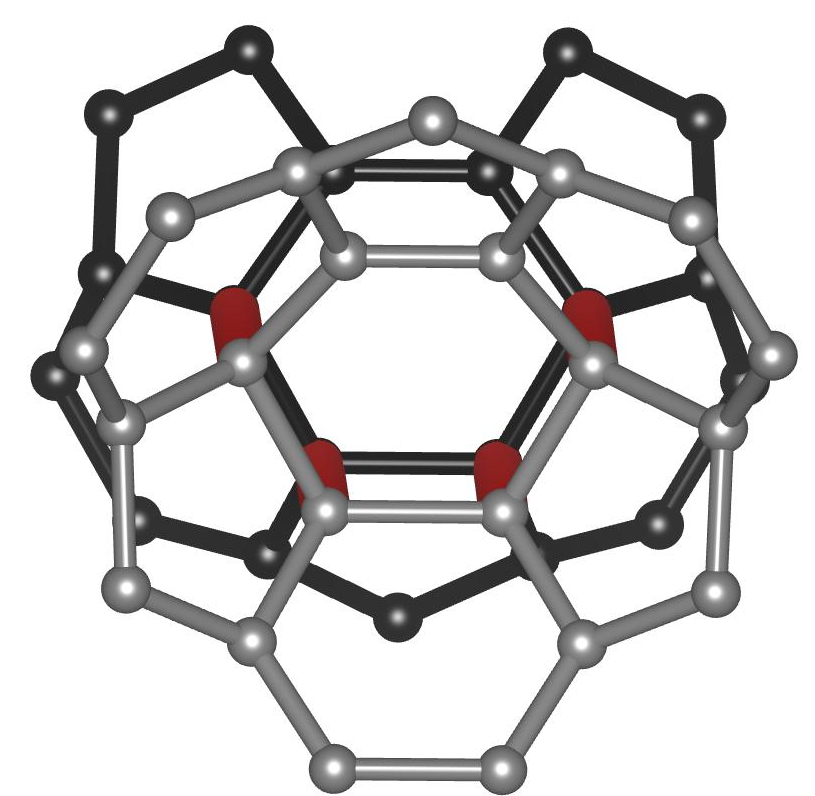}  & AMK1269 \\ 
	   &  &    &      \\
	   \hline
	   	   &    &  &      \\
	   Min8	5/66 3+2  & 2 &  \includegraphics[scale=0.1]{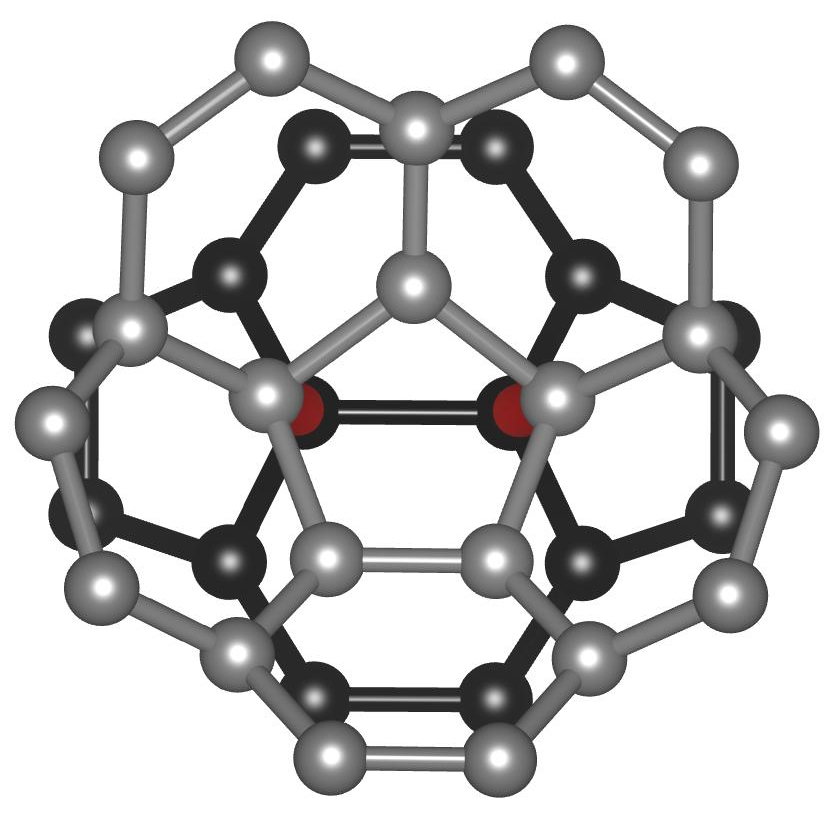}  & AMK3333 \\
	   &    &  &      \\
	    \hline
	    &  &    &      \\
	   Min9	double 56/56 2+2 & 4 &  \includegraphics[scale=0.1]{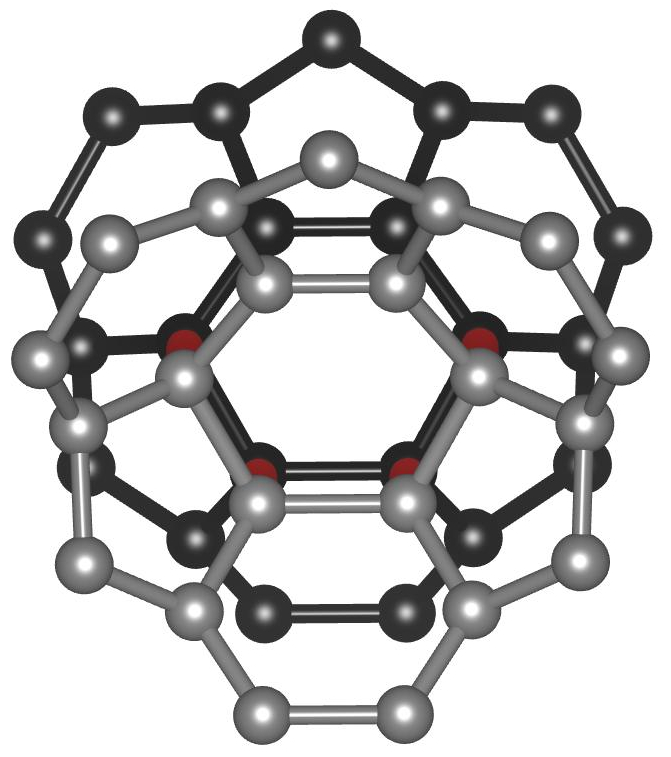} & G \\
	   &  &    &      \\
	   \hline
	   &    &  &      \\
	   Min10 AMK5678   &  3 & \includegraphics[scale=0.1]{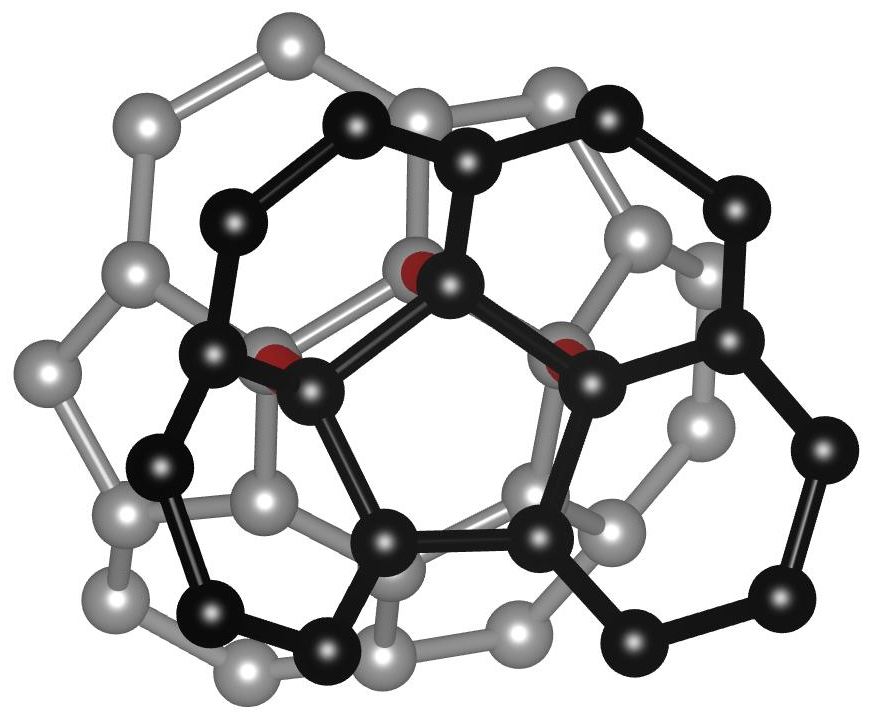}  & AMK5678  \\
&    &  &      \\
\hline
&  &  &       \\
Min11 6/66 3+2 &  2 &    \includegraphics[scale=0.1]{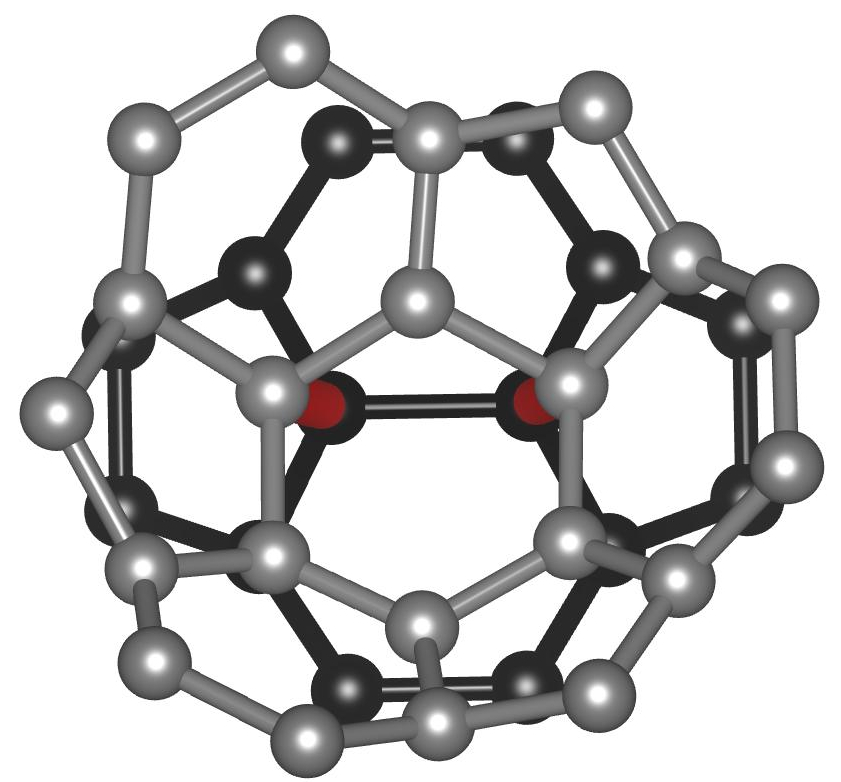} &  AMK3615 \\
&  &  &       \\
\hline
&    &  &      \\
Min 12 	6/6 4+4 M1 &  2 &  \includegraphics[scale=0.1]{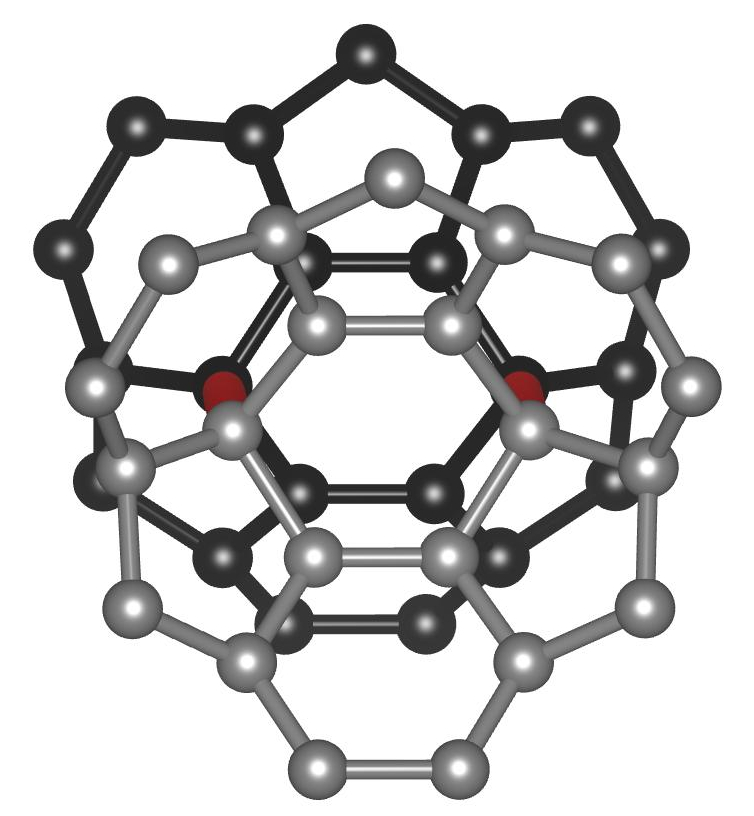}   & AMK2506/L \cite{osawa_dimeros,Matsuzawa_dimeros} \\  %$\mathrm{C}_{2\mathrm{v}}$ ?
&    &  &      \\
\hline 
 &  &  &      \\
Min13	triple 66/65 2+2  & 6 & \includegraphics[scale=0.1]{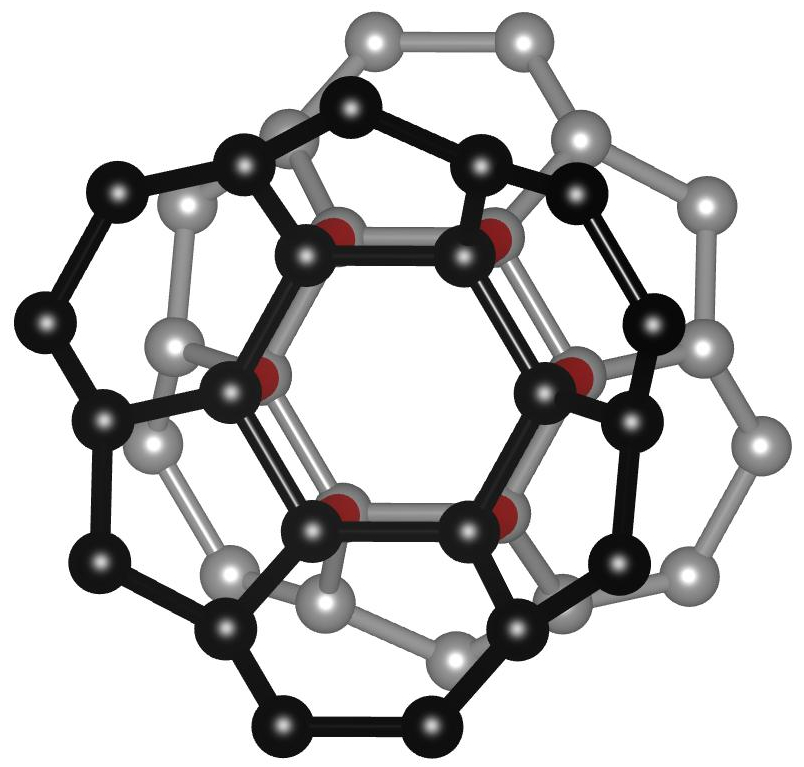}  &    AMK1372/L  \cite{adamsprb94_dimeros,LIU2007_dimeros}\\
&  &  &      \\
\hline
&    &  &      \\
Min 14	6/6 4+4 M2  & 2 & \includegraphics[scale=0.1]{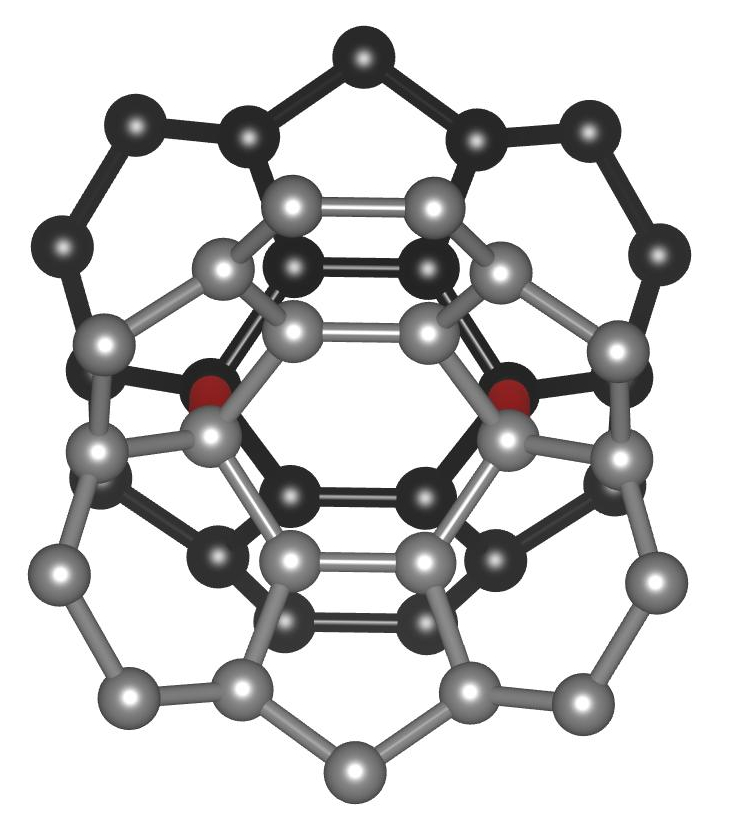}   &  AMK2507/L \cite{osawa_dimeros,Matsuzawa_dimeros}  \\ %$\mathrm{C}_{2\mathrm{h}}$ ?
&  &  &        \\
\hline
&  &  &      \\
Min15	triple 66/66 2+2  & 6 & \includegraphics[scale=0.1]{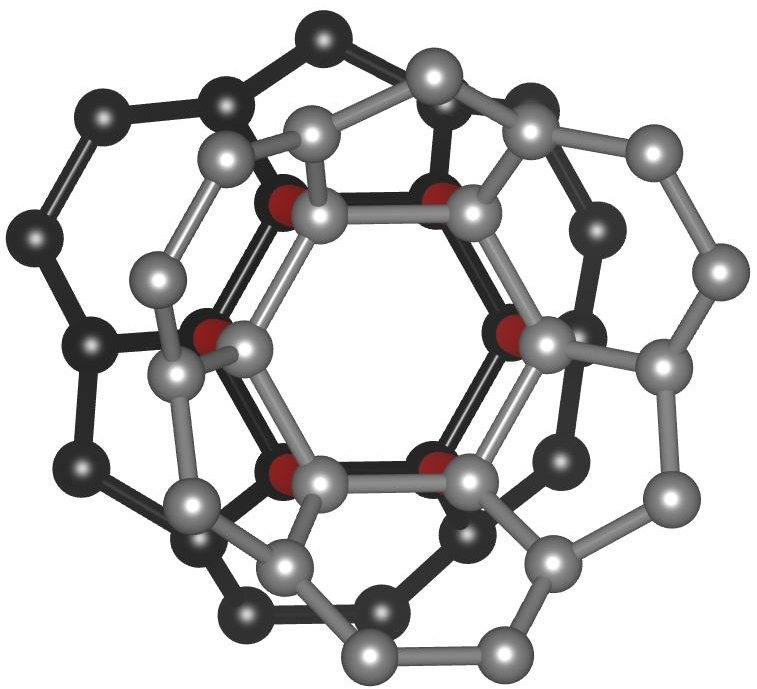}    &    L  \cite{adamsprb94_dimeros,LIU2007_dimeros}\\
&  &  &      \\
 \hline
&    &  &      \\
Min 16 	6/6 3+4 M2   & 2 & \includegraphics[scale=0.1]{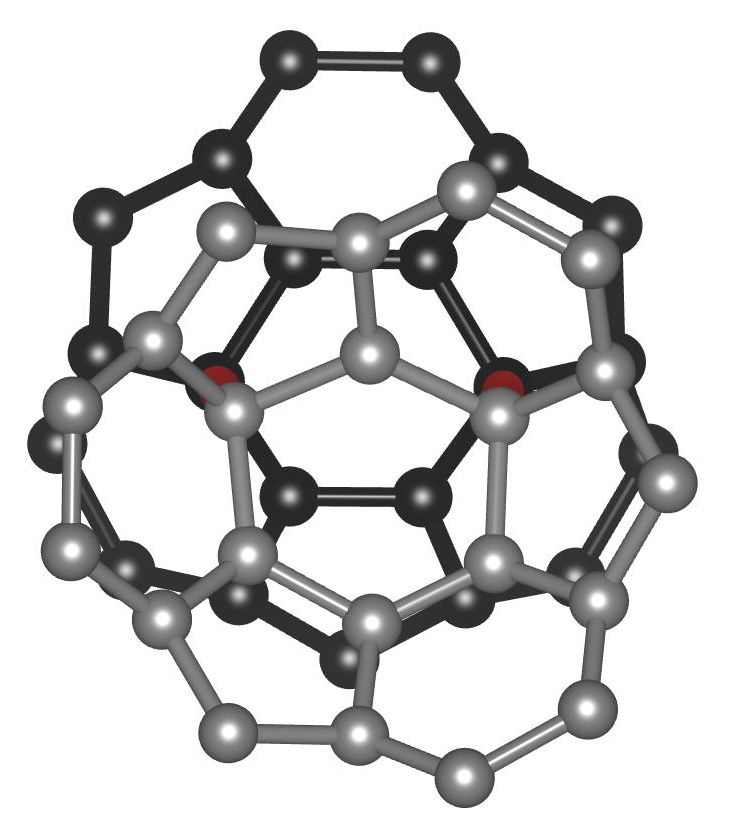}    & AMK4004 \\
&    &  &      \\
\hline
&    &  &      \\
Min 17 6/6 3+3 M3   & 2 & \includegraphics[scale=0.1]{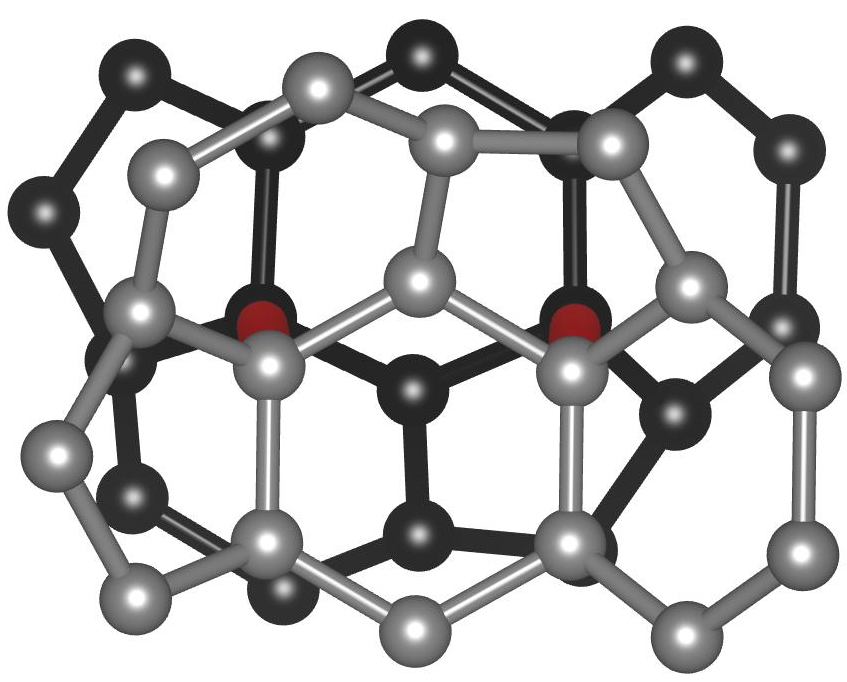}   & AMK5325  \\
&    &  &      \\
\hline
&  &  &      \\
Min18	double 56/66 2+2  M1 & 4 & \includegraphics[scale=0.1]{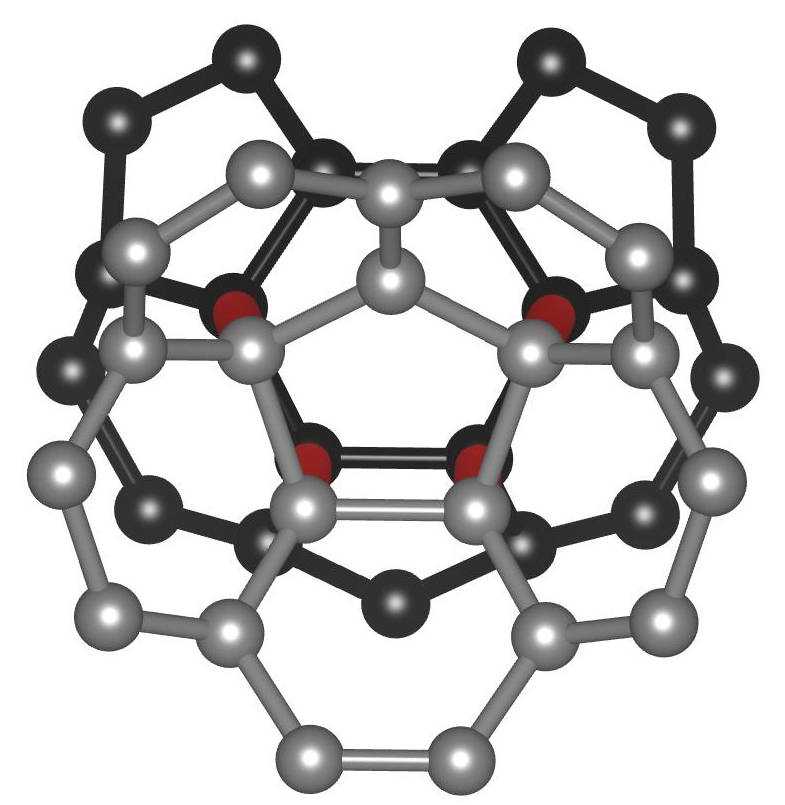}   & AMK2605\\ 
&    &  &      \\
	\hline
&  &  &        \\
Min 19	5/6 3+4 M2  & 2 & \includegraphics[scale=0.1]{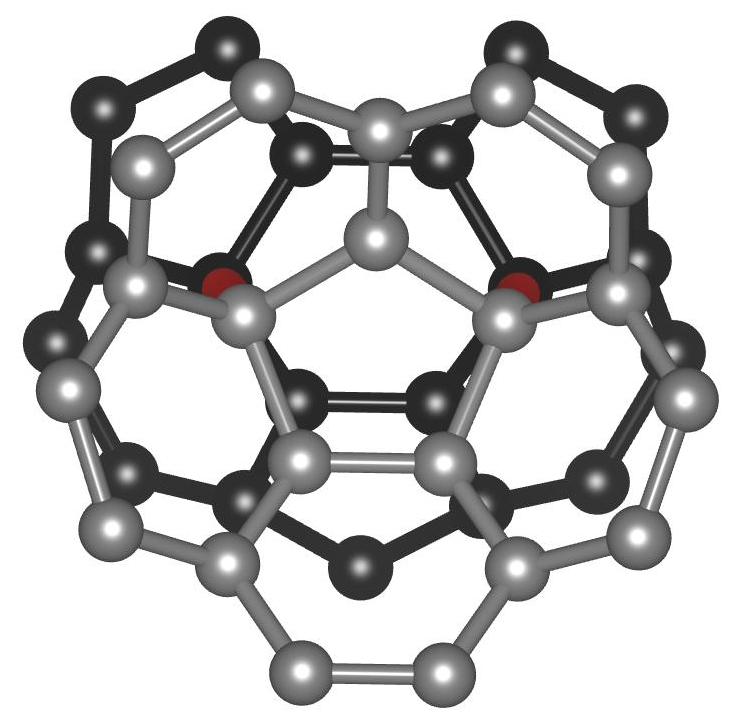}  &   AMK4307 \\
&  &  &        \\
\hline
&  &  &        \\
Min 20	5/6 3+4 M1  & 2 & \includegraphics[scale=0.1]{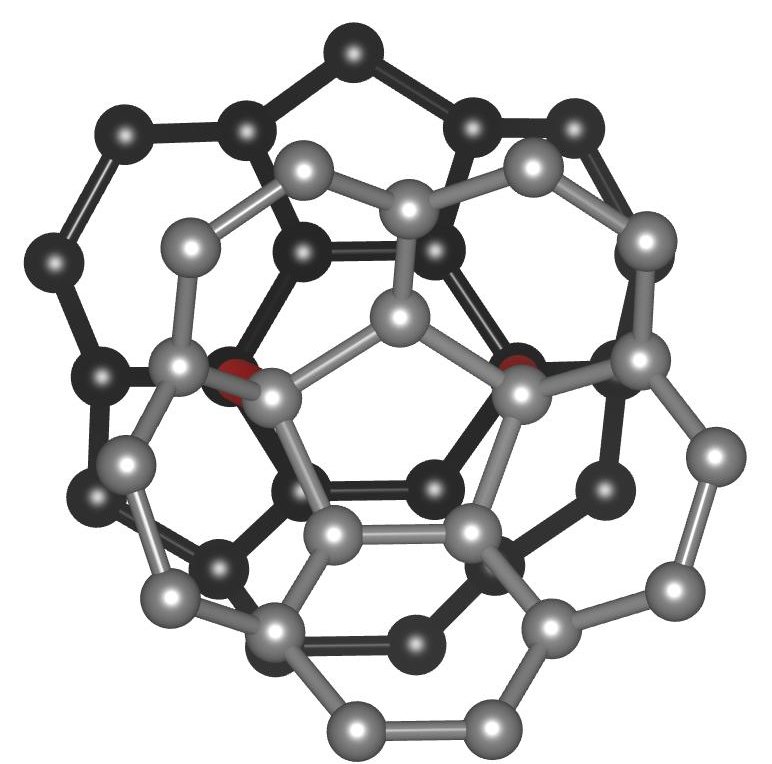}  &   AMK4321 \\
&  &  &        \\
\hline
&  &  &        \\
Min 21 6/56 3+2 M2 &  2 &  \includegraphics[scale=0.1]{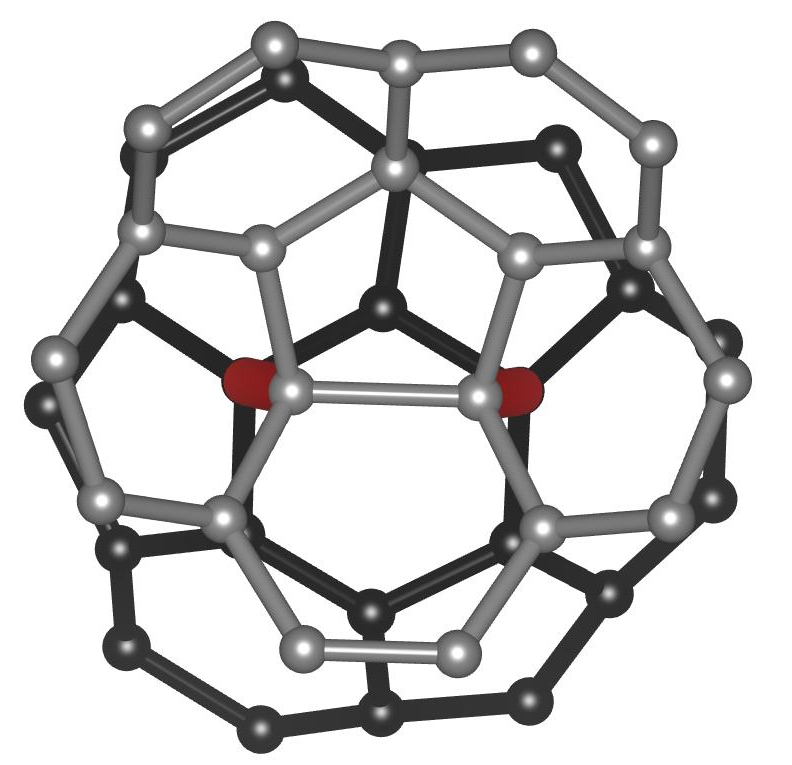}   &  AMK4646 \\
&  &  &        \\
\hline
&  &  &       \\
Min22 6/56 3+2 M1 &  2 & \includegraphics[scale=0.1]{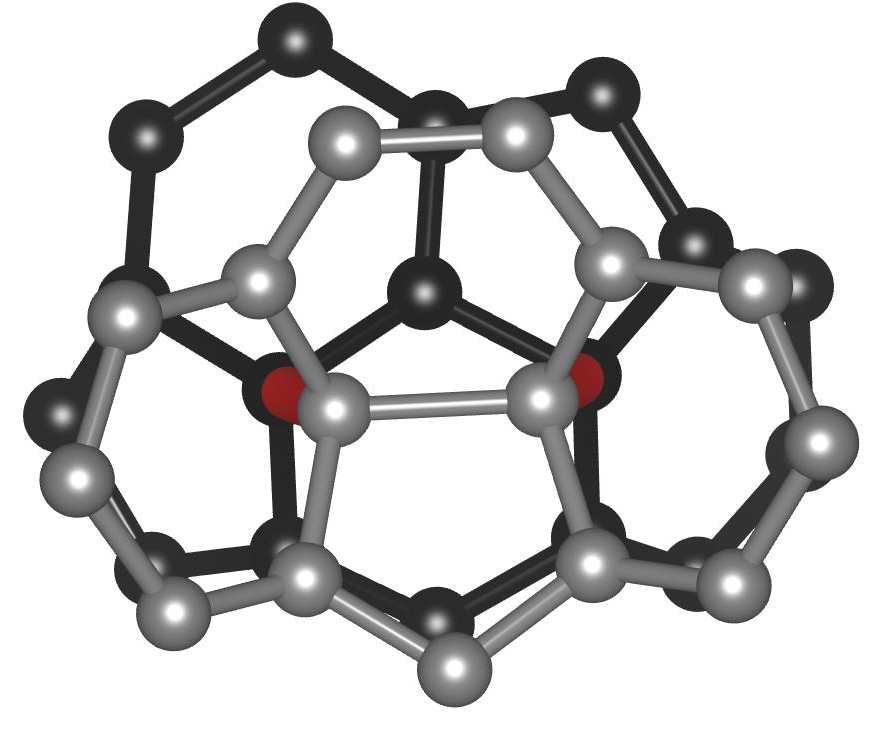}   & AMK4567  \\
&  &  &        \\
\hline
&    &  &      \\
Min23	double 56/65 2+2   & 4 & \includegraphics[scale=0.1]{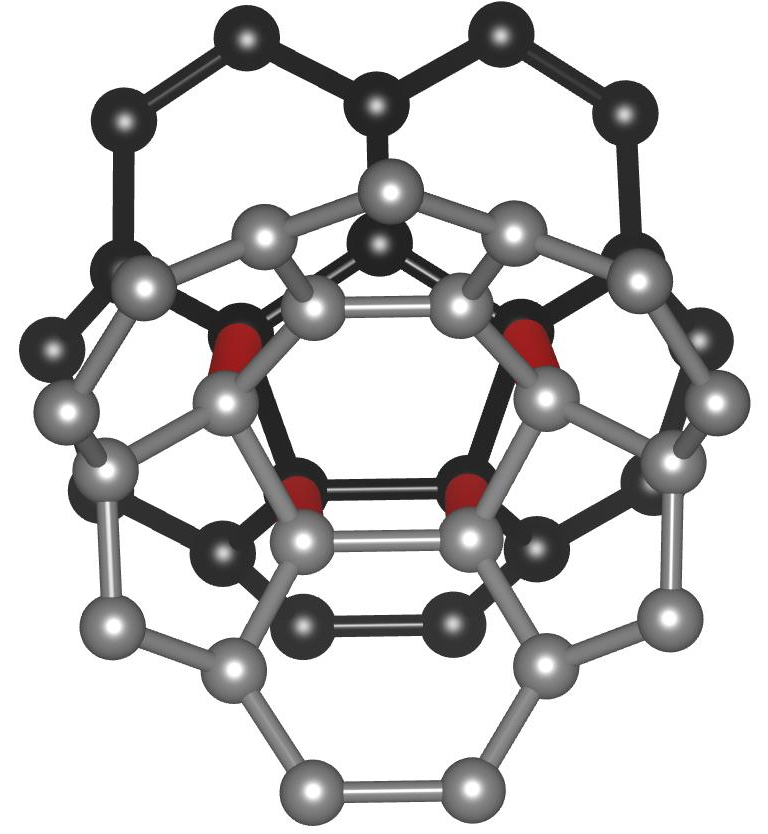}  & AMK3018 \\
&    &  &      \\
 \hline
 	 &    &  &      \\
 Min 24 5/5 3+3 M1 &  2 &    \includegraphics[scale=0.1]{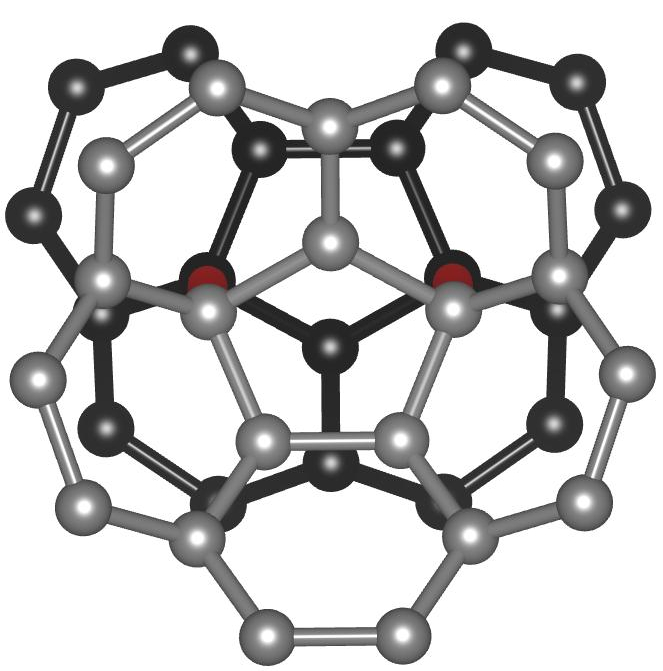}   & G \\
 &  &  &      \\
 \hline
 &    &  &      \\
 Min25 6/6 3+4 M1   & 2 &  \includegraphics[scale=0.1]{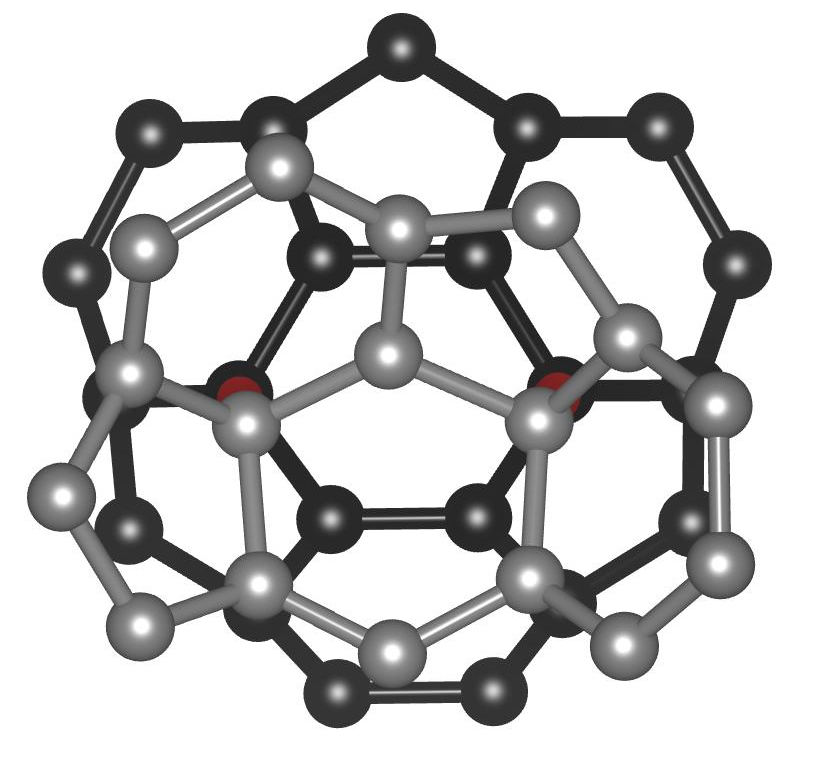}   & AMK3893  \\
 &    &  &      \\
 \hline
 &    &  &      \\
 Min 26 	double 56/56 2+2   & 4 & \includegraphics[scale=0.1]{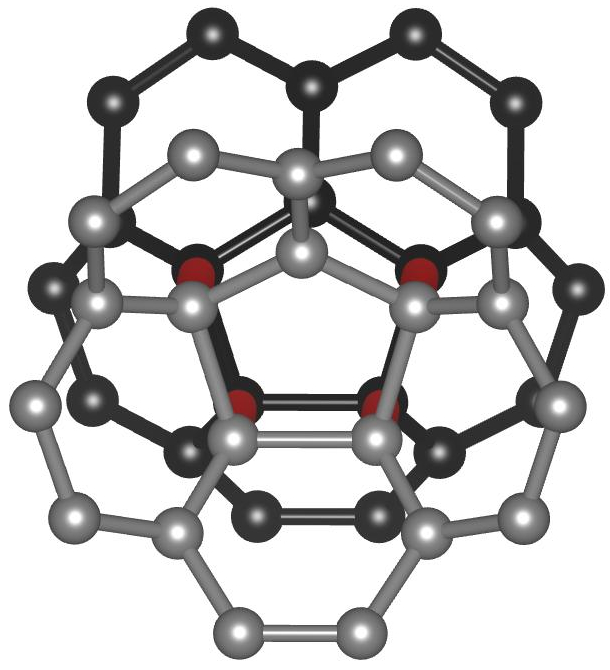}   &    G \\
 &    &  &      \\
 	\hline
 &    &  &      \\
 Min 27 AMK6163  & 3 &  \includegraphics[scale=0.1]{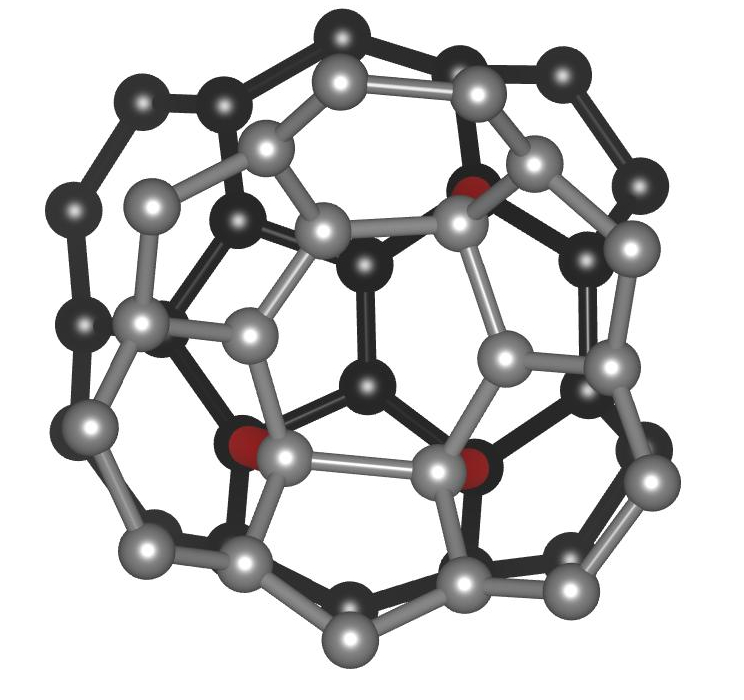} &    AMK6163 \\
  \hline
 	   &  &  &      \\
 Min 28 double 56/56 2+2 plus SB  & 5 &  \includegraphics[scale=0.1]{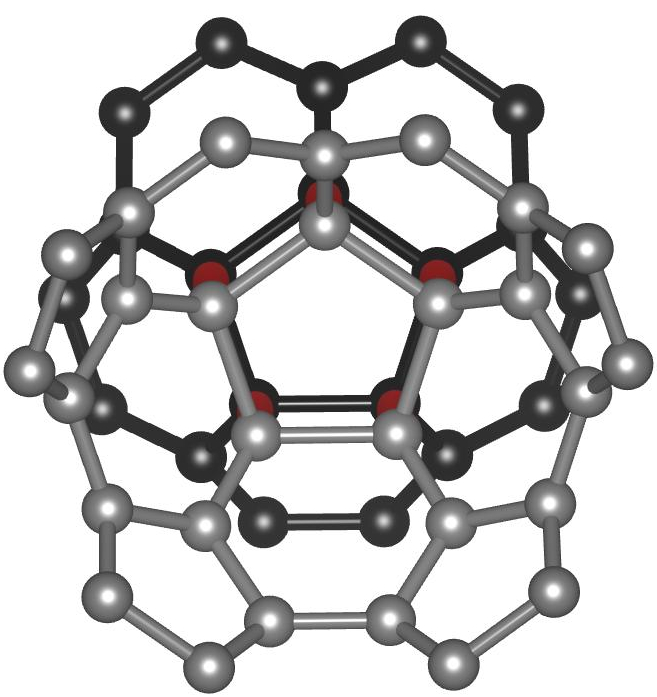} & L \cite{adamsprb94_dimeros}\\
 &  &    &      \\
 \hline
 &  &  &        \\
 Min 29 	double 6/56 4+2 &  4 & \includegraphics[scale=0.1]{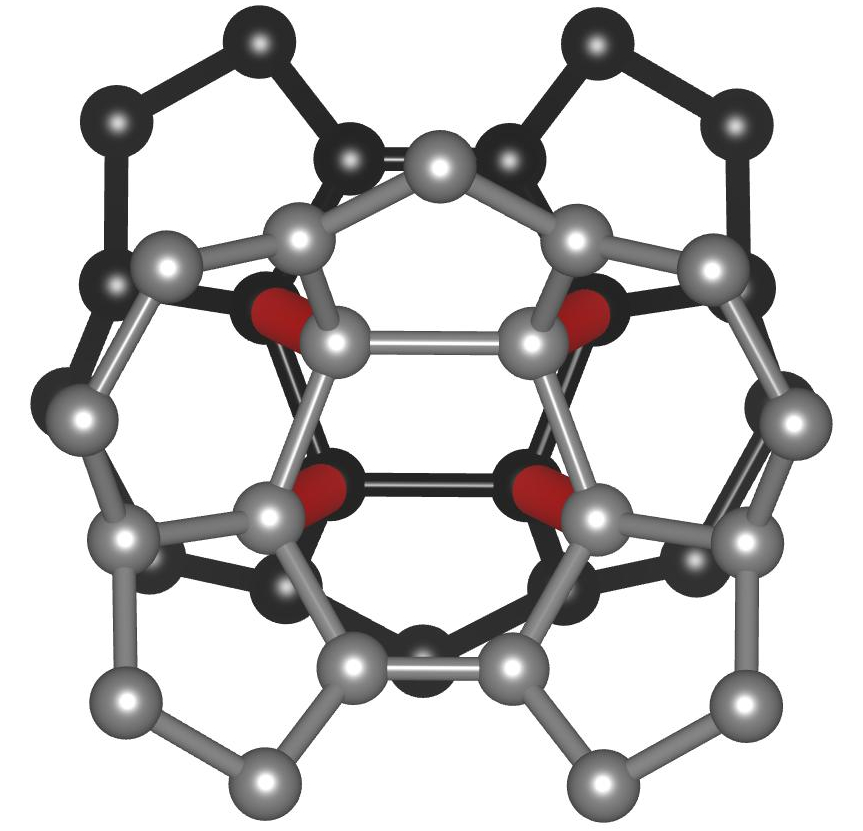}   & L \cite{strout1993dim} \\
 &    &  &      \\
  \hline
 &    &  &      \\
 Min 30 5/56 3+2  &  &      &  \\
 
 plus  &  &      & G \\
 
 6/56 4+2  & 4 &     \includegraphics[scale=0.1]{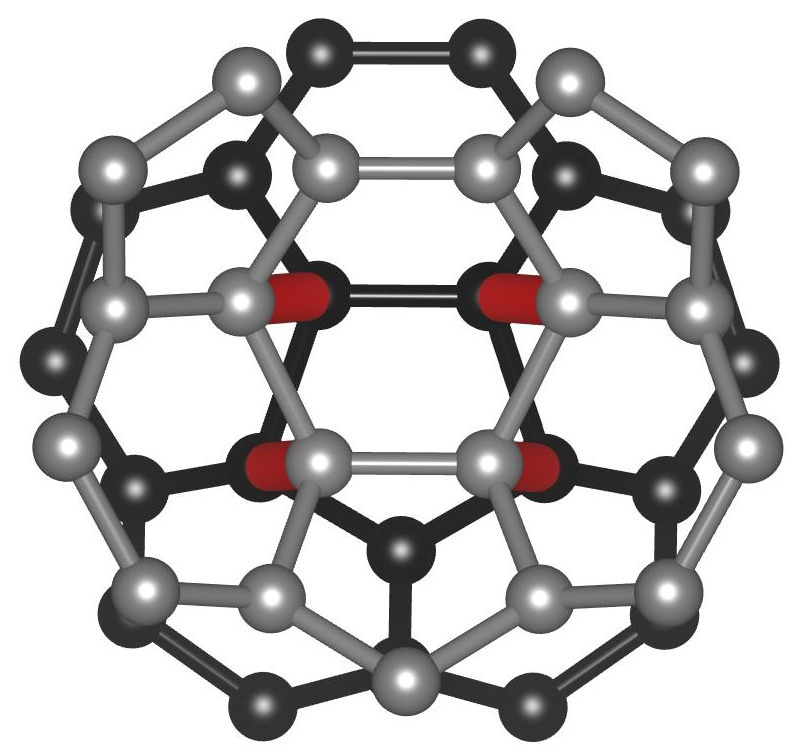}  &  \\
 &  &    &      \\
  \hline
    &  &  &      \\
  Min 31	double 5/5 2+3 &  4 & 	\includegraphics[scale=0.1]{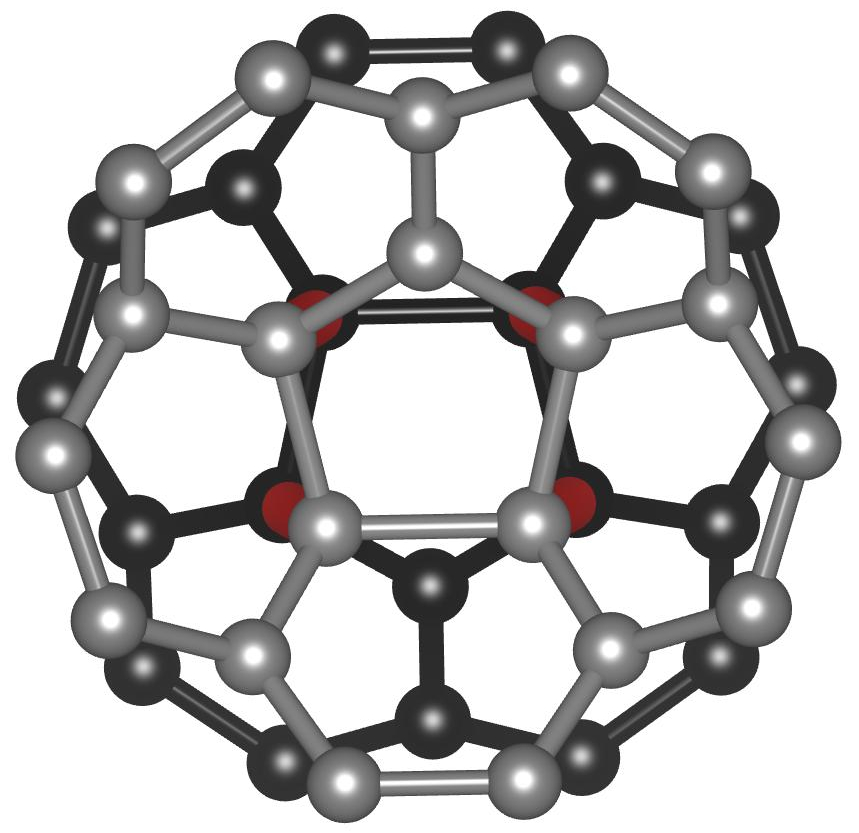} & G \\
  &    &  &      \\
  \hline
 &  &  &        \\
 Min 32 5/6 3+3 plus 56/6 2+3  & 4 & \includegraphics[scale=0.1]{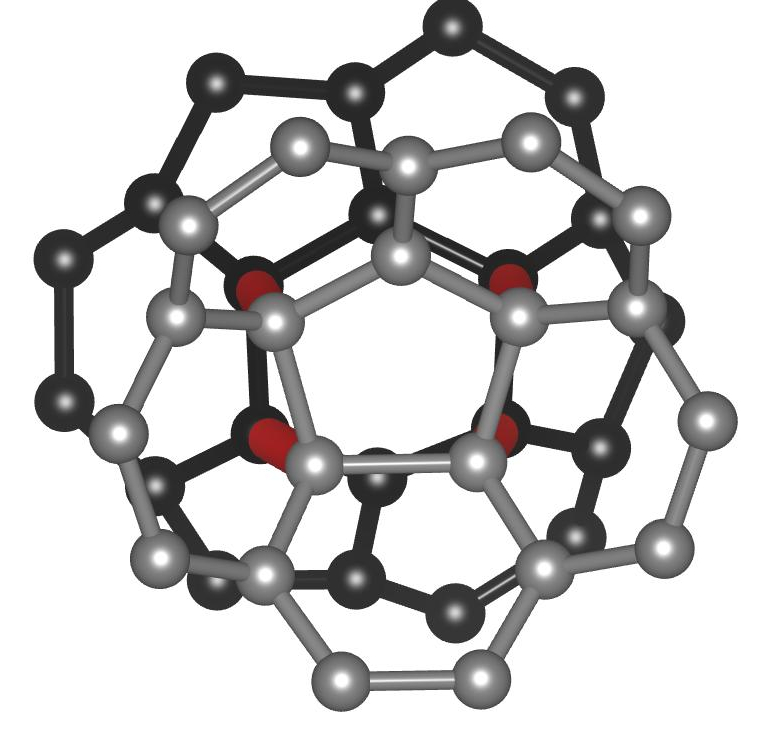}  &    AMK5775 \\
 &  &    &      \\
 \hline
 &  &    &      \\
 Min 33	AMK7077  & 4 & \includegraphics[scale=0.1]{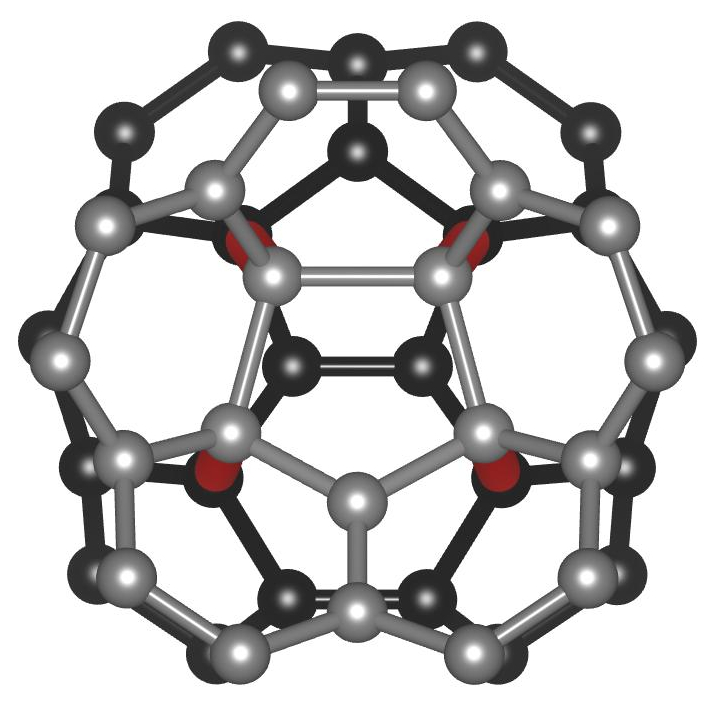} &     AMK7077 \\
 &  &    &      \\
 	\hline
 &    &  &      \\
 Min 34 AMK6511  & 4 & \includegraphics[scale=0.1]{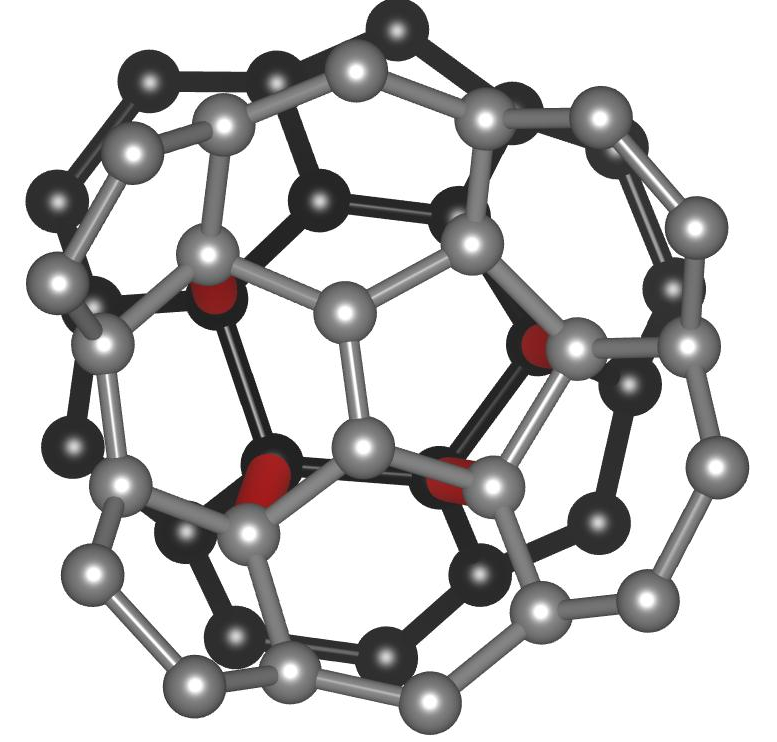}  &     AMK6511 \\
 &  &  &      \\
 \hline
 &  &  &       \\
 Min35 AMK7610 &  4 & \includegraphics[scale=0.1]{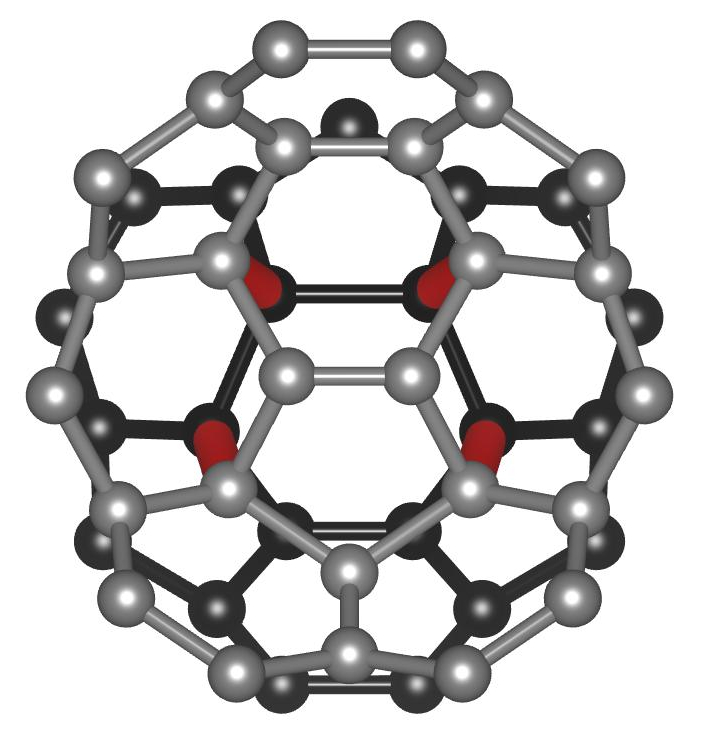}  &    AMK7610 \\
 &  &    &      \\
 \hline
 &  &    &      \\
 Min36	5/5 3+3 plus 6/6 4+4  & 4 & \includegraphics[scale=0.1]{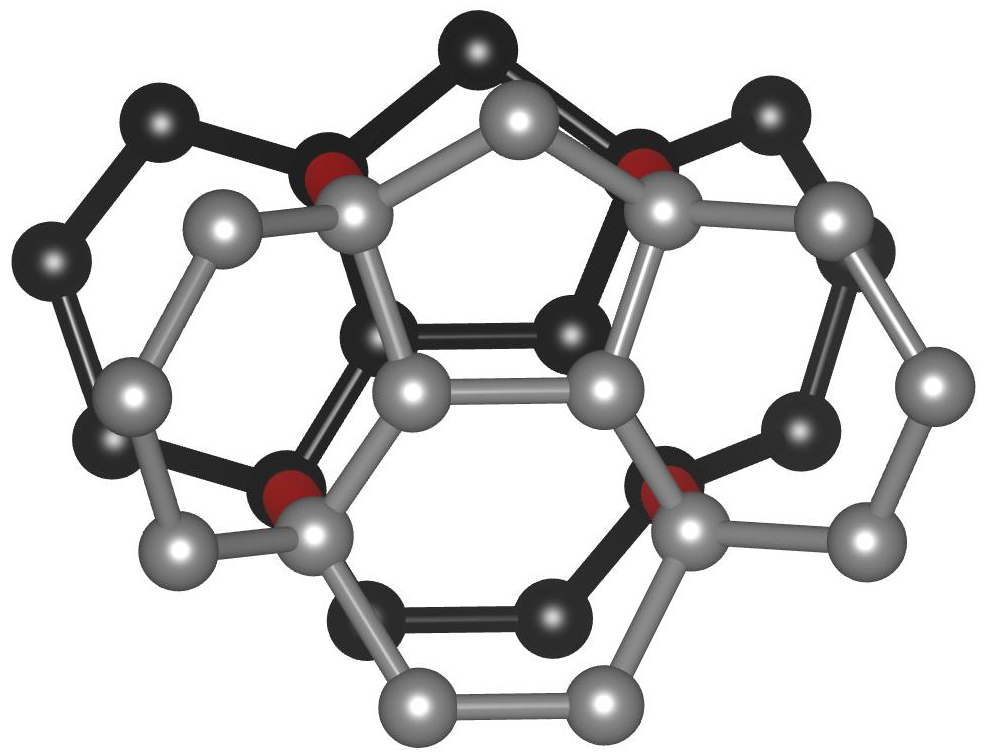}  &     G \\
 &    &  &      \\
 \hline
  &  &  &        \\
 Min37	double 56/65 3+4  & 4 & \includegraphics[scale=0.1]{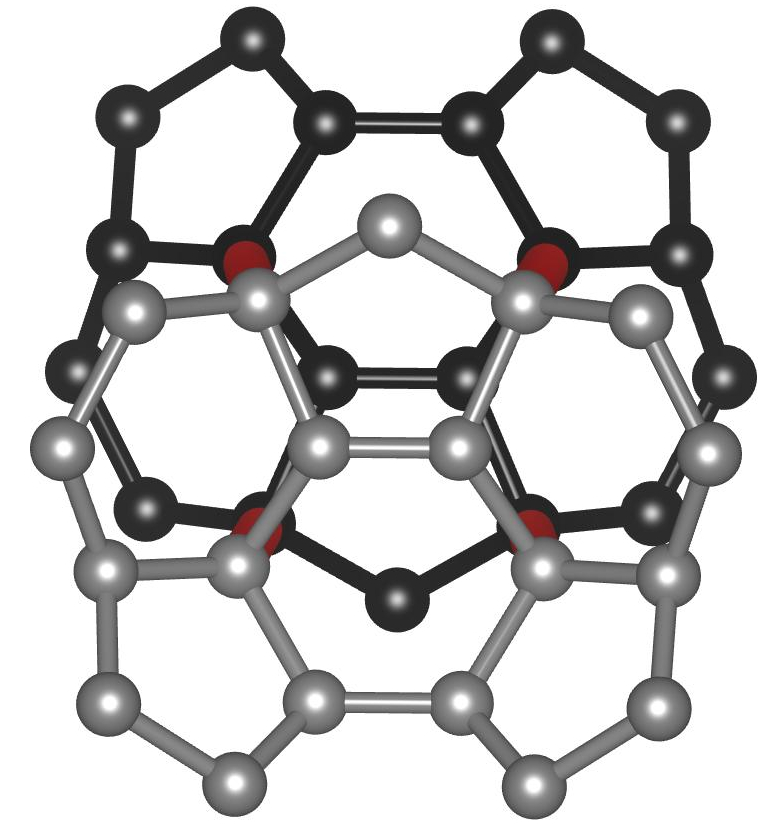}  &     S-AMK\_or  \\%5563
 &    &  &      \\
  \hline
 &  &   &      \\
 Min38	5/6 3+4 plus 6/6 4+4   & 4 & \includegraphics[scale=0.1]{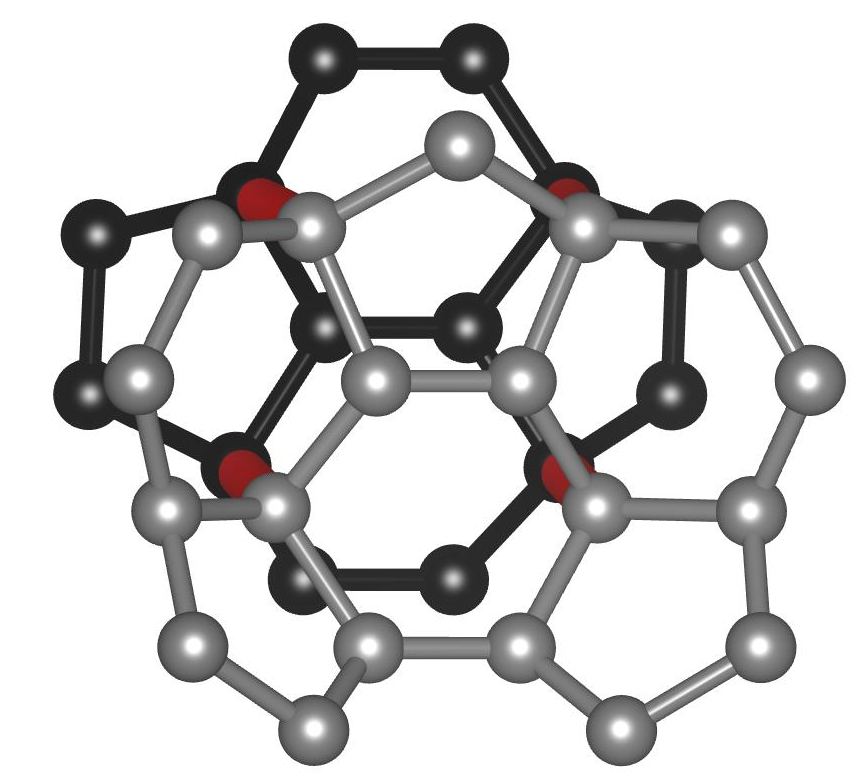}  &    G \\
 &  &    &      \\
 \hline
 &  &  &        \\
 Min39	double 66/66 4+4  & 4 & \includegraphics[scale=0.1]{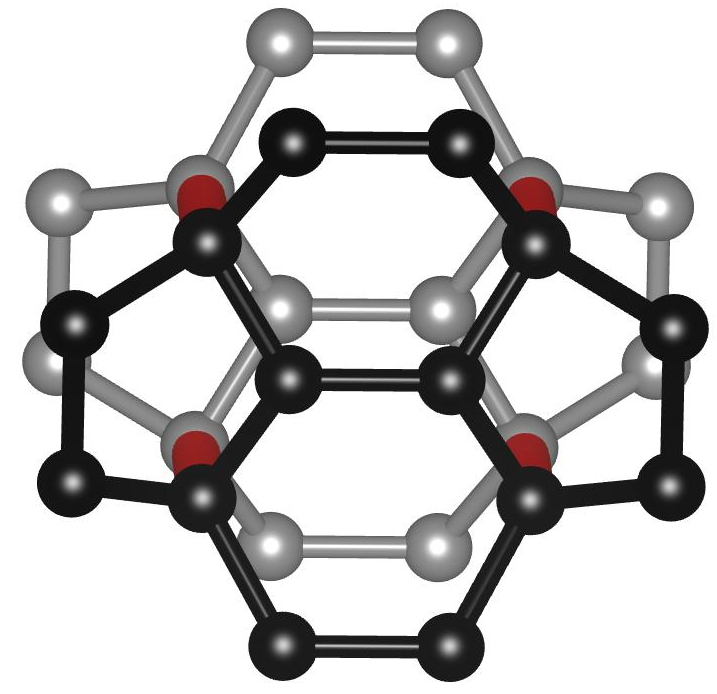}  &     G \\
 &  &    &      \\
 \hline
  &  &  &        \\
 Min 40 AMK8161  & 4 &  \includegraphics[scale=0.1]{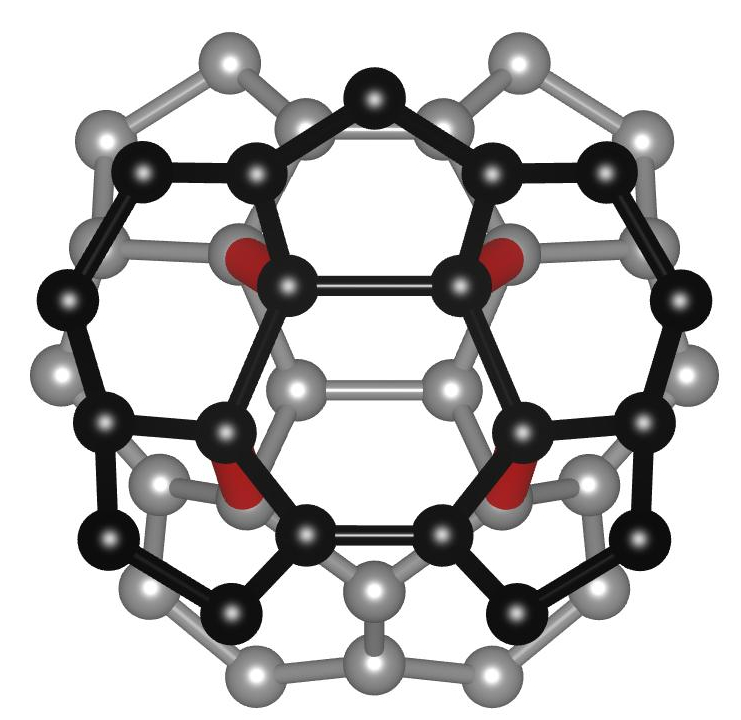}  &     AMK8161 \\
 &  &    &      \\
 \hline
 &  &  &        \\
 Min41	double 66/66 4+4    &  &  &     \\
 plus     & 6 & \includegraphics[scale=0.1]{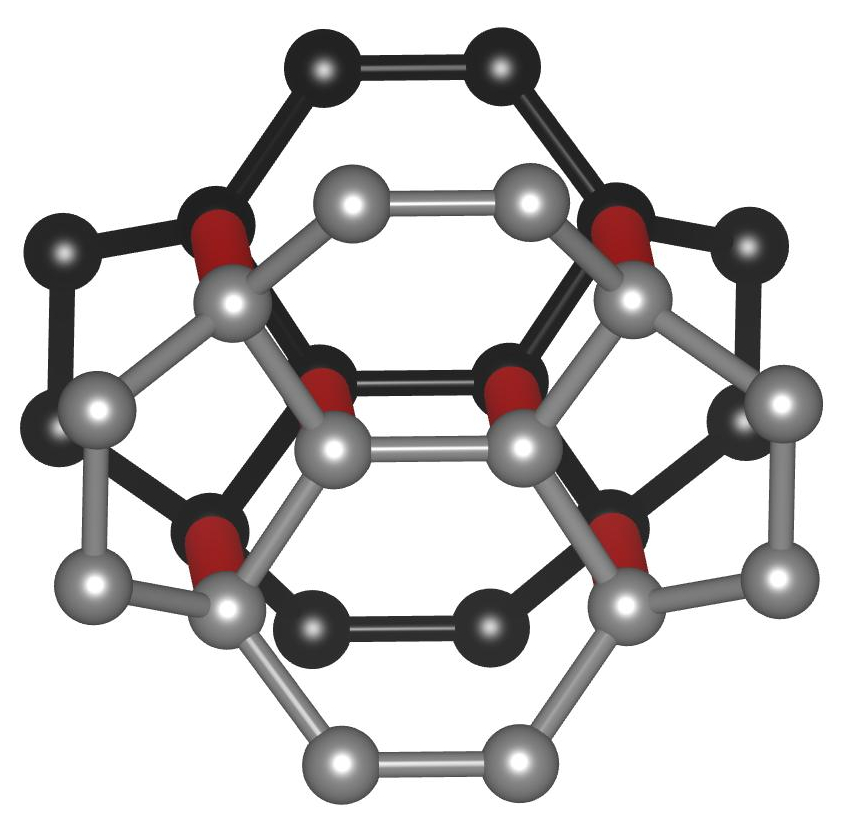}  &  G \\
 
 66/66 2+2   &  &  &   \\
 &  &  &      \\
 \hline
 &  &  &   \\
	double 66/6 2+4    & 4 & \includegraphics[scale=0.1]{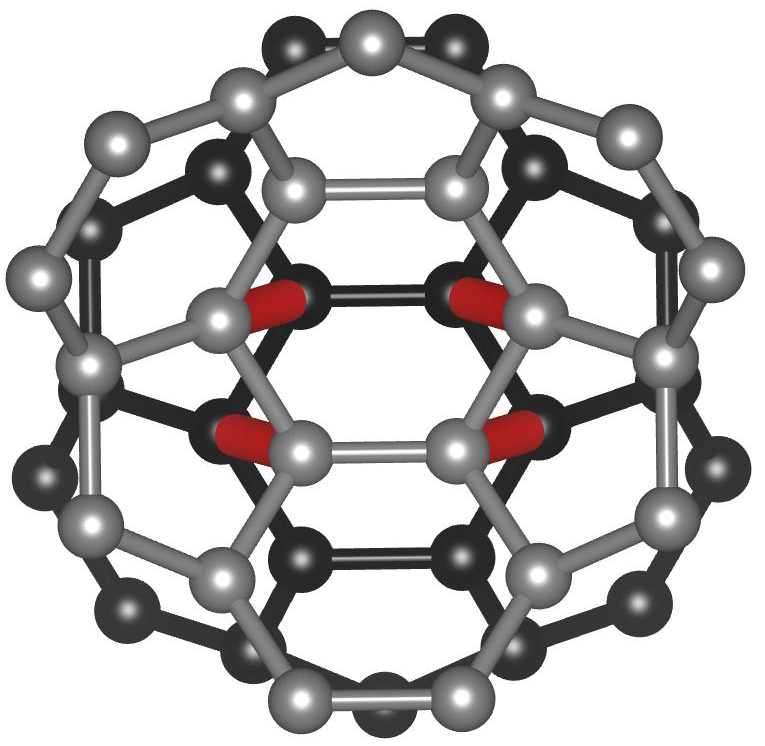}  &     G \\ %qual dos de simetria é\\
	 &  &  &   \\
\hline
  &  &  &   \\
	66/6 2+4   & 2 & \includegraphics[scale=0.1]{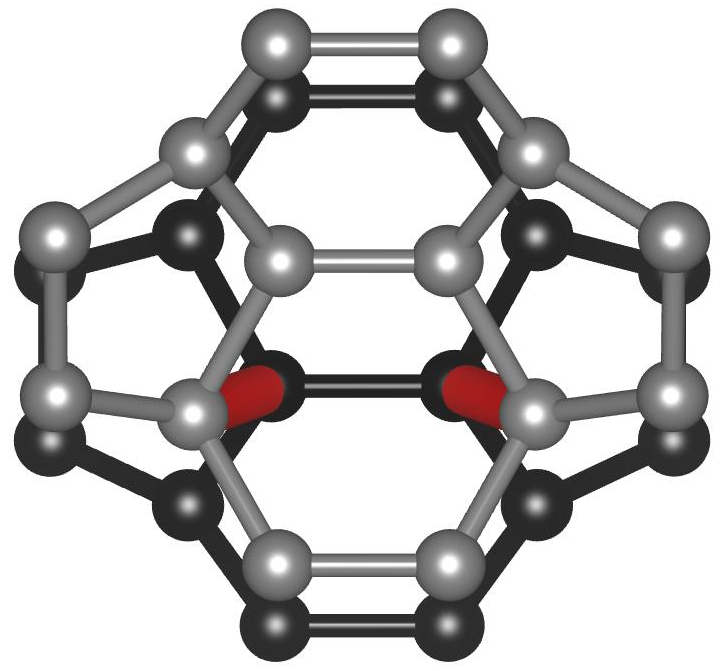}  &   AMK3097/L  \cite{Matsuzawa_dimeros} \\%dos de simetria é\\
  &  &  &   \\
\hline
	   &  &  &      \\
	5/5 3+3 M2  & 2 &     \includegraphics[scale=0.1]{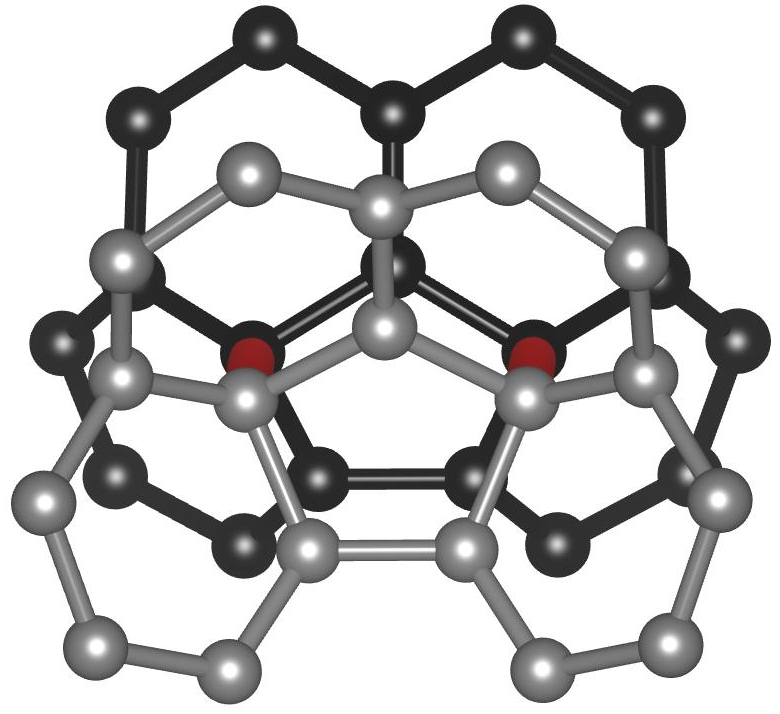}  & G \\
		   &  &  &      \\
\hline
	 &  &    &      \\
	5/56 3+2  & 2 &    \includegraphics[scale=0.1]{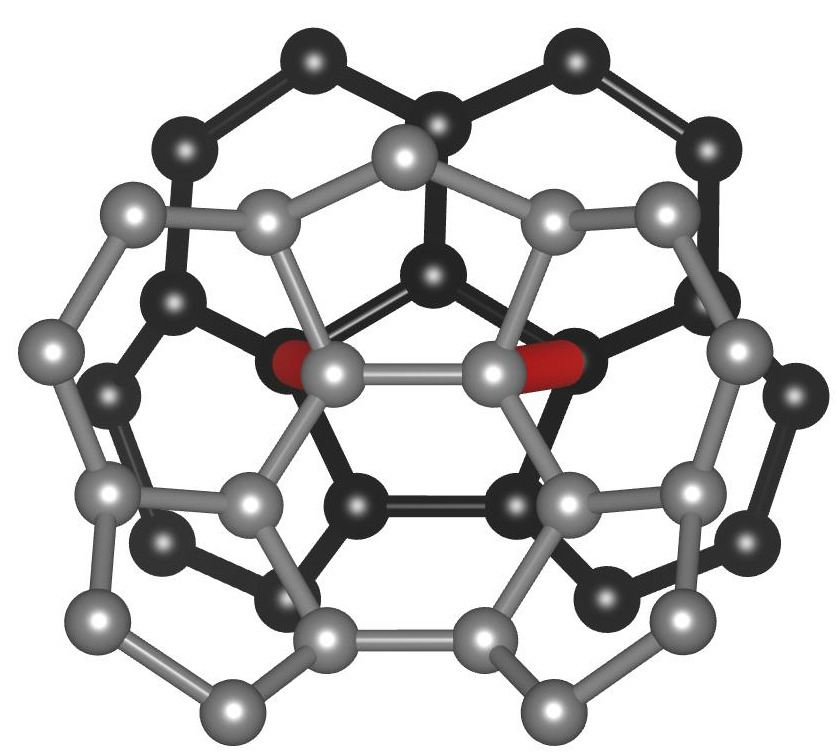}  & G \\
	 &    &  &      \\
\hline
 &    &  &      \\
	5/65 3+2  & 2 &  \includegraphics[scale=0.1]{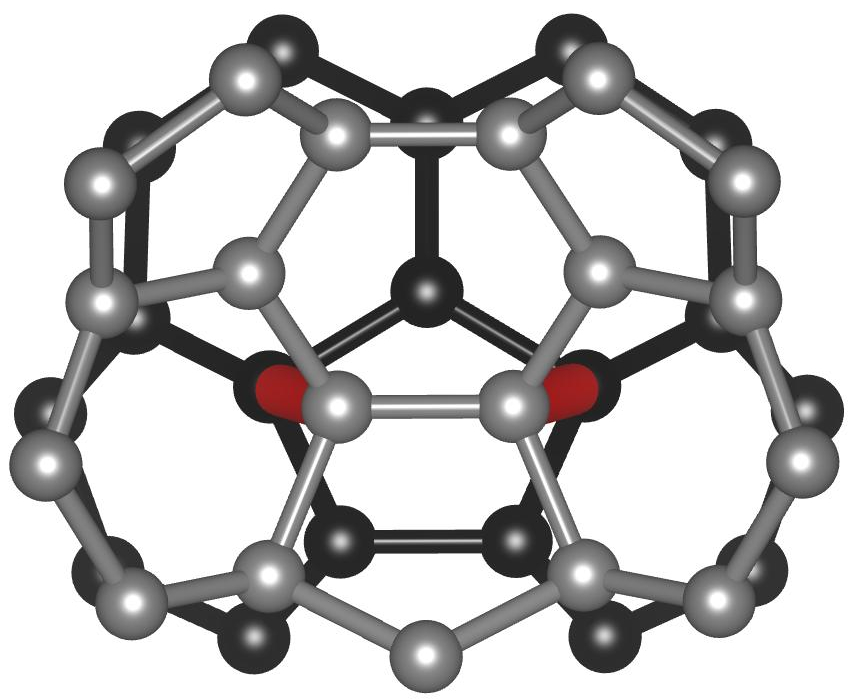}   & G \\
	 &  &    &      \\
\hline
 &  &    &      \\
	6/6 3+3 M1 &   2 &   \includegraphics[scale=0.1]{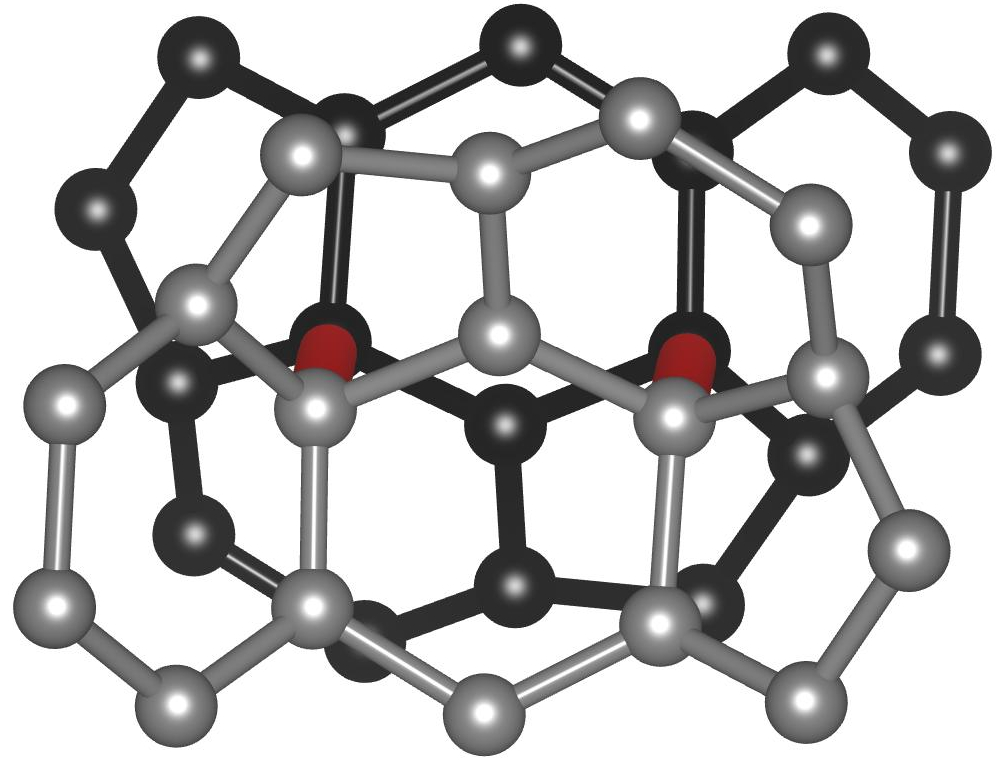}    & G  \\
	 &   &  &      \\
\hline
	 &    &  &      \\
	6/6 3+3 M2  & 2 &    \includegraphics[scale=0.1]{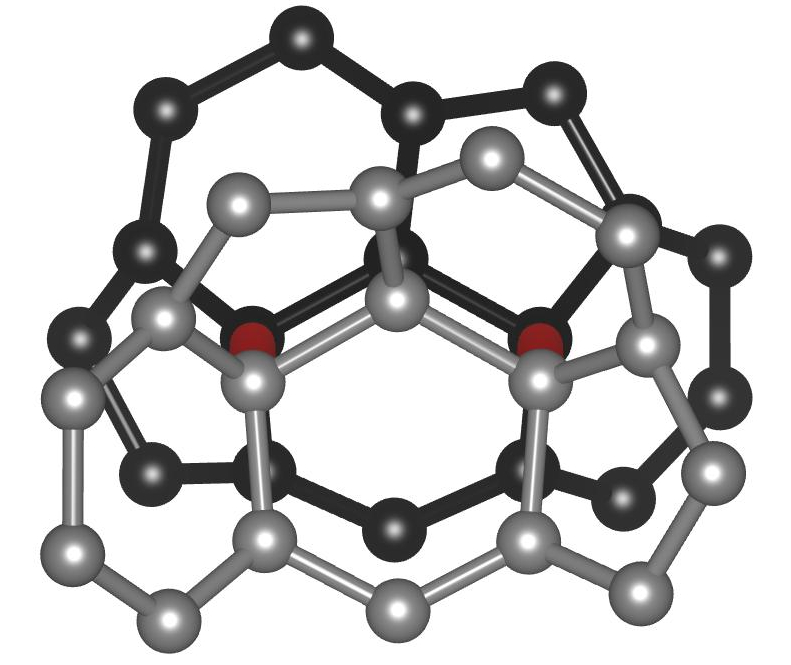}   & AMK5282  \\
		 &    &  &      \\
\hline
 &    &  &      \\
	6/6 3+3 M4   & 2 & \includegraphics[scale=0.1]{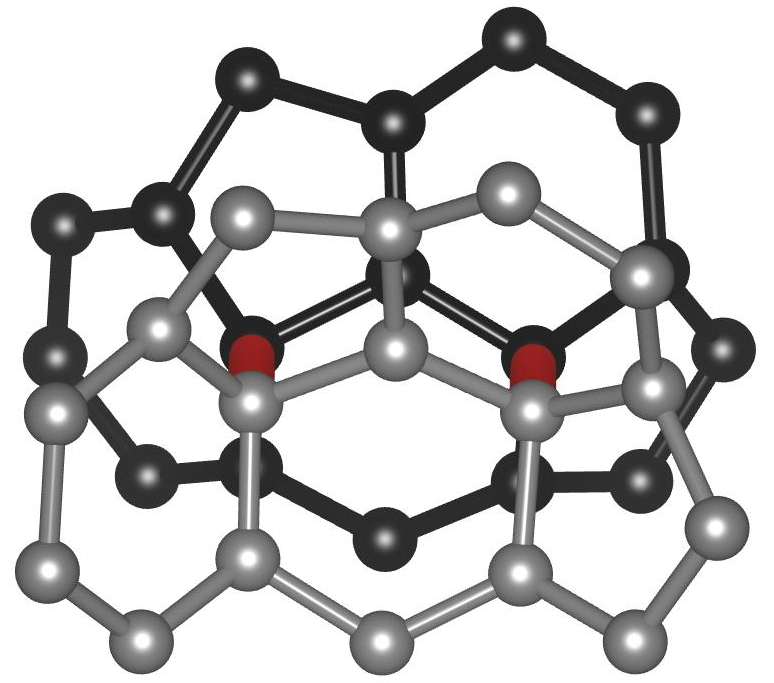}    & G  \\
 &    &  &      \\
	\hline
		 &  &  &        \\
	6/56 4+2 M1 &  2 &    \includegraphics[scale=0.1]{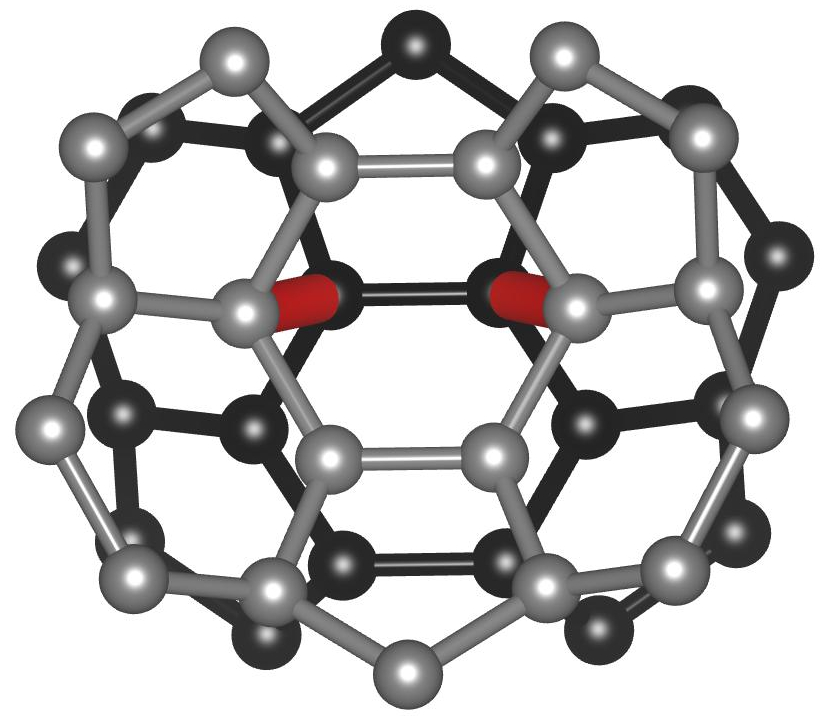}   & G  \\
		 &  &  &        \\
	\hline
		 &  &  &        \\
	6/56 4+2 M2  & 2 &    \includegraphics[scale=0.1]{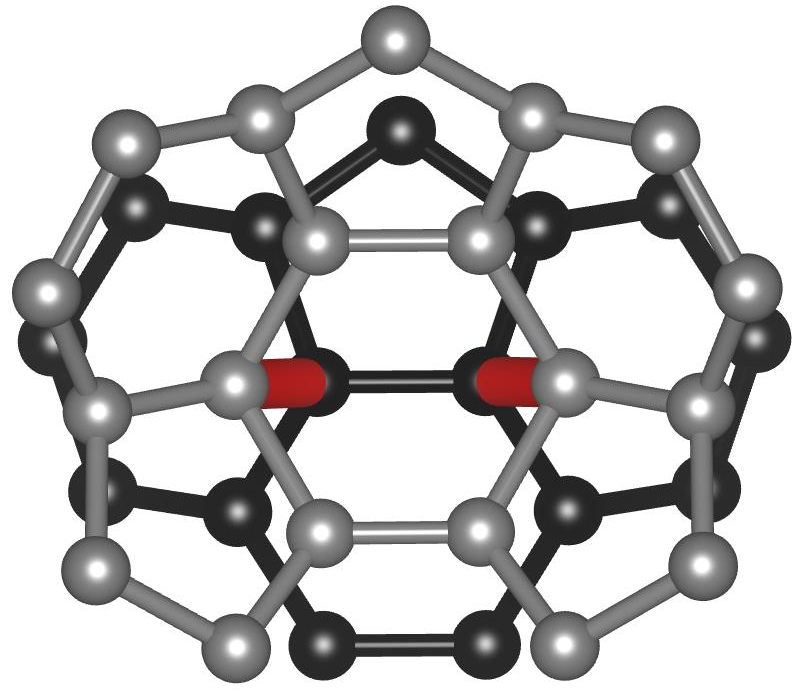}   &  G \\
	 &  &  &        \\
\hline
	 &  &  &        \\
	6/56 4+2 plus 6/66 4+2   & 4 &  \includegraphics[scale=0.1]{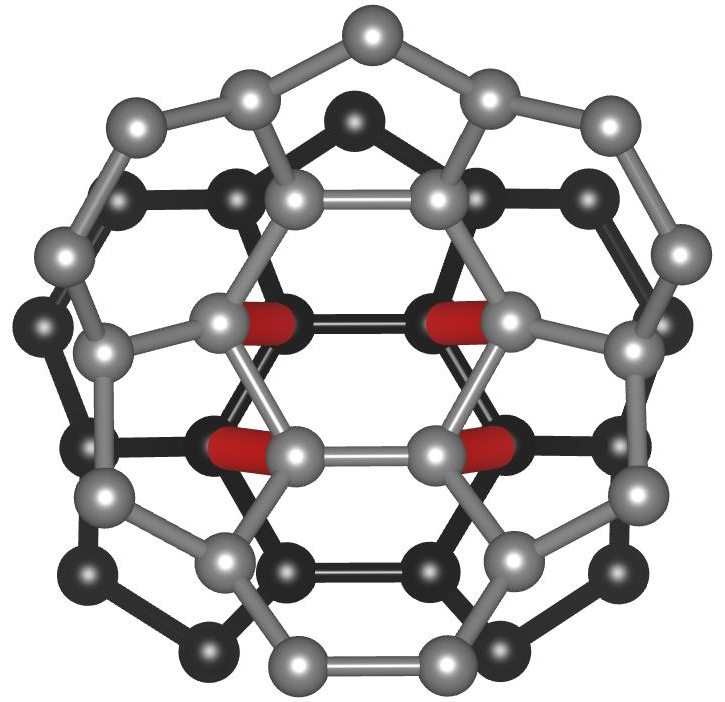}   & G  \\
	 &  &  &       \\
	\hline
	 &  &   &      \\
	5/66 3+2 plus 6/56 4+2  & 4 &    \includegraphics[scale=0.1]{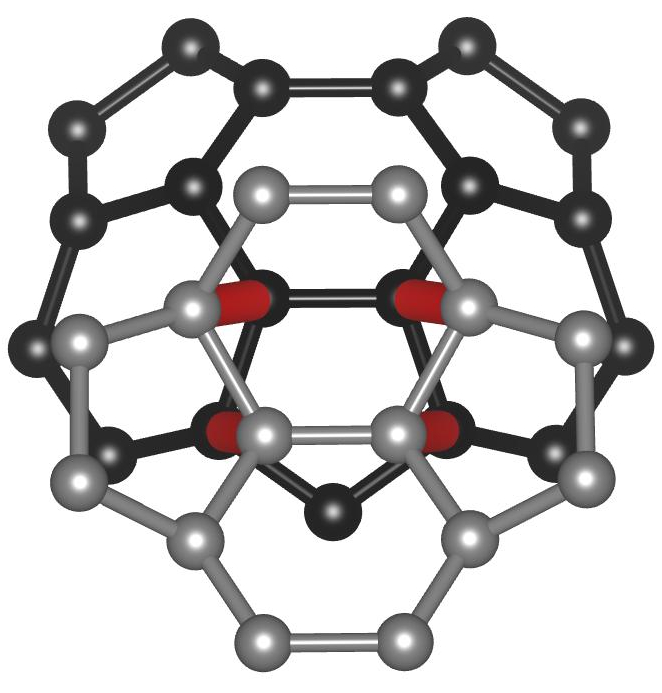}  & G \\
	 &  &  &        \\
	 	\hline
	 	 &  &  &       \\
	5/6 3+3 M1 & 2 &   \includegraphics[scale=0.1]{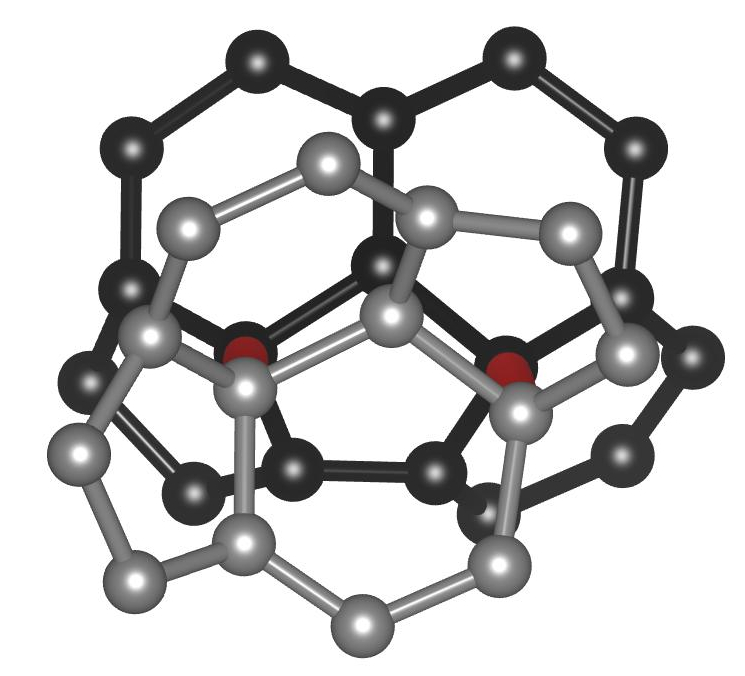}  & AMK5118\\ %??AMK-5678 \\
		 &  &  &        \\
	\hline
	&  &  &        \\
	%zigzag
	double 66/66 4+4 zig   & 4 & \includegraphics[scale=0.1]{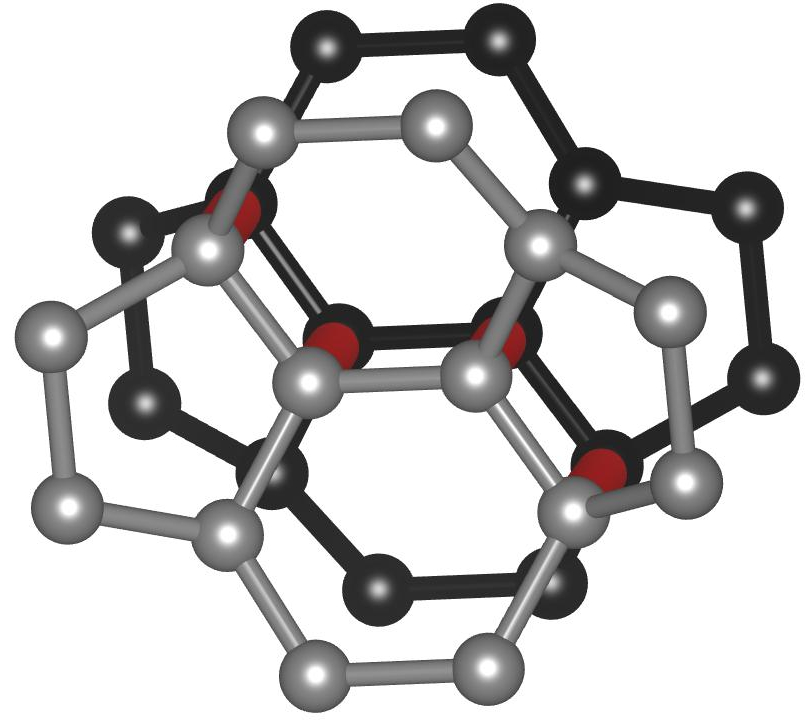} &  G \\
			 &  &  &        \\
	\hline
	&  &  &        \\
	double 56/65 3+4  zig   & 4 &\includegraphics[scale=0.1]{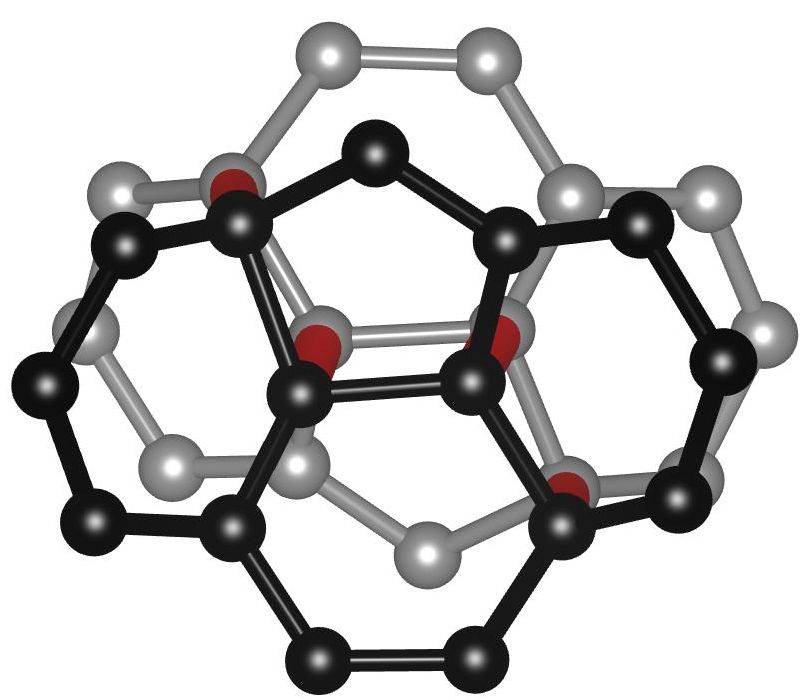} &  G \\
			 &  &  &        \\
	\hline
	&  &  &        \\
	5/5 3+3 plus 6/6 4+4  zig   & 4 & \includegraphics[scale=0.1]{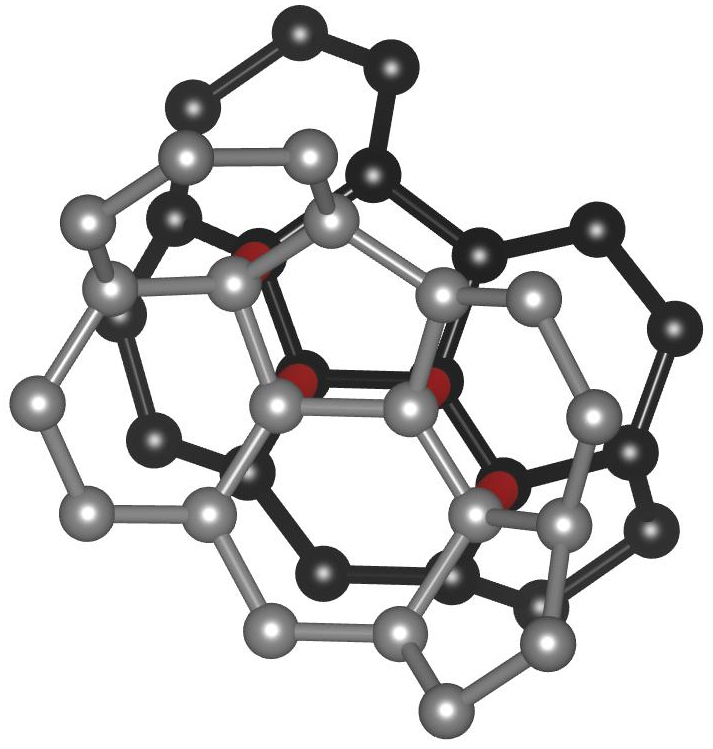} &   G \\
%	5/6 3+4 plus 6/6 4+4  &  & 4 &  &   G \\
	%
		 &  &  &        \\
\hline
%	double 66/66 4+4  &  &  &  &     \\
%	plus 66/66 2+2 plus  &  & 4 &  &   G \\
%	double 56/56 2+2 &  &  &  &   \\
%	\hline
&  &    &      \\
	double 66/66 4+4   & &\includegraphics[scale=0.1]{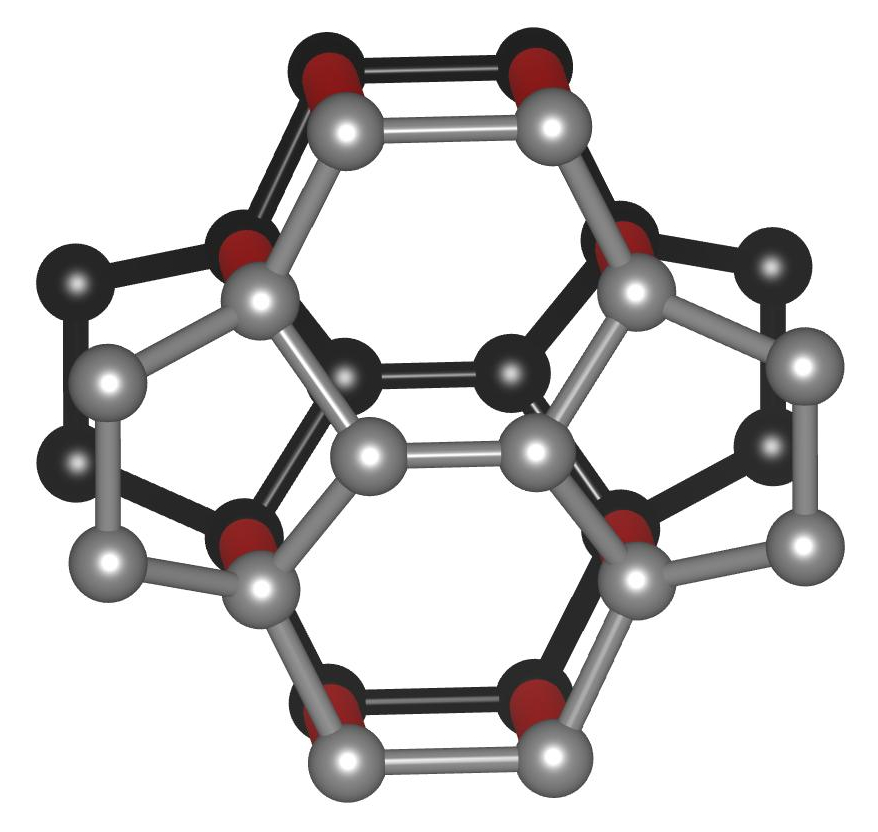}     &     \\
	plus    & 4 &  &   G \\
	double 56/56 2+2   &  &  &   \\
		   &  &  &      \\
\hline
&  &    &      \\
	AMK5282 &   3 &  \includegraphics[scale=0.1]{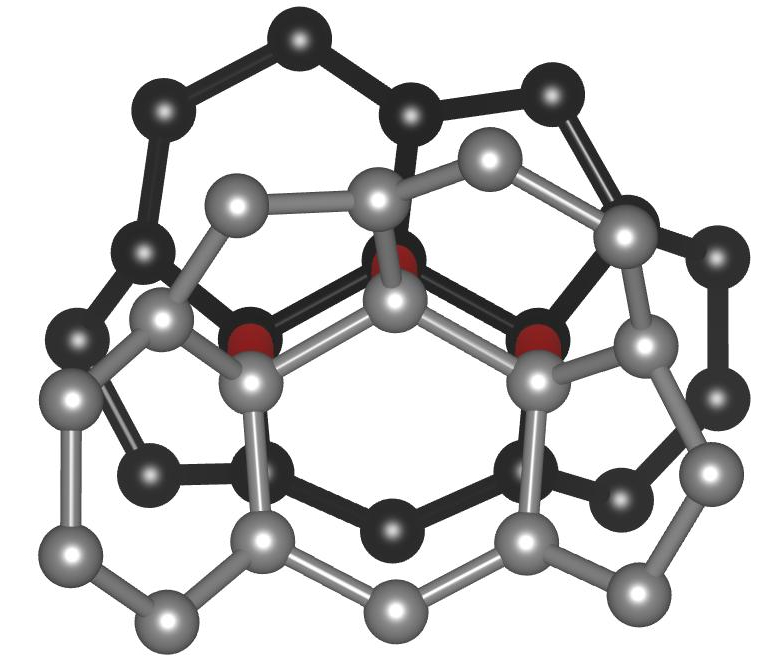} & \\  % AMK5282??? \\
			   &  &  &      \\
	\hline
	&    &  &      \\
	AMK6480   & 4 & \includegraphics[scale=0.1]{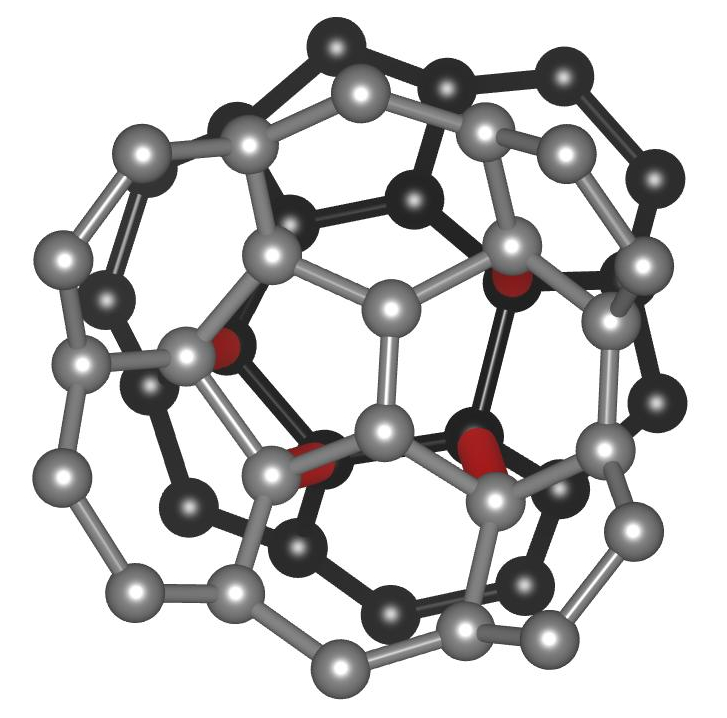} &    AMK6480 \\
	%%	5/6 3+3 M2 & & 2 & \includegraphics[scale=0.1]{jpgs1/.jpg}  &   S-AMK6993 \\	
	
	%	\end{tabular}
	
	\label{at_apendix}
\end{longtable}

\bibliography{referencias2.bib}